\documentclass[a4paper,11pt]{article}
\pdfoutput=1 % if your are submitting a pdflatex (i.e. if you have
\usepackage{jcappub} % for details on the use of the package, please
\usepackage[T1]{fontenc} % if needed
\usepackage{bm}
\usepackage{mathtools}
\usepackage[
backend=bibtex,
eprint=true,
doi=false,
isbn=false,
url=false]{biblatex}
\title{\boldmath Statistics of biased tracers in variance-suppressed simulations}

\author[a,b]{Francisco Maion}
\author[a,c]{, Raul E. Angulo}
\author[a]{, Matteo Zennaro}

\affiliation[a]{Donostia International Physics Center (DIPC), Paseo Manuel de Lardizabal, 4, 20018, Donostia-San Sebastián, Guipuzkoa, Spain}
\affiliation[b]{Departamento de Física Matemática, Instituto de Física,
Universidade de São Paulo, Rua do Matão 1371, CEP 05508-090, São Paulo, Brazil}
\affiliation[c]{IKERBASQUE, Basque Foundation for Science, 48013, Bilbao, Spain.}

\emailAdd{francisco.maion@dipc.org}
\emailAdd{reangulo@dipc.org}
\emailAdd{matteo\_zennaro001@ehu.eus}

\abstract{Cosmological simulations play an increasingly important role in analysing the
observed large-scale structure of the Universe. Recently, they have been
particularly important in building hybrid models that combine a perturbative
bias expansion with displacement fields extracted from N-body
simulations to describe the clustering of biased tracers.  Here, we show that simulations that employ a technique referred
to as "Fixing-and-pairing" (F\&P) can dramatically improve the statistical
precision of such hybrid models.  Specifically, by numerical and analytical
means, we show that F\&P simulations provide unbiased estimates for all
statistics employed by hybrid models while reducing, by up to two orders of
magnitude, their uncertainty on large scales.  This roughly implies that an
EUCLID-like survey could be analysed using simulations of 2Gpc a side -- a 20\%
of the survey volume. Our work establishes the robustness of F\&P for current
hybrid theoretical models for galaxy clustering, an important step towards
achieving an optimal exploitation of large-scale structure measurements.
}

\begin{document}
\maketitle
\flushbottom

\section{Introduction}

Numerical simulations of the large-scale structure are fundamental tools in modern cosmology \cite{angulo_hahn}. They can be used for a variety of purposes such as generating mock observations of the universe, testing the underlying physics of the cosmos and assessing the observability of novel effects, or even to generate large training sets for machine-learning based models that seek to describe complex phenomena. Finally, they are the only tools capable of accurately describing the nonlinear regime of structure formation.

However, many of the aforementioned applications suffer from a limitation -- these numerical simulations poorly sample modes comparable with the fundamental mode of the simulated volume. This issue is related to the observational effect known as \textit{cosmic variance}, and causes a lack of precision in statistical quantities on large scales. It is possible to increase this precision by running simulations with larger volumes or many realizations of them, but of course there are limits imposed by the capabilities of even the largest supercomputers available. No simulation run up to now is capable of matching the combination of volume and resolution required for the interpretation of data from Stage IV galaxy surveys such as EUCLID or Rubin-DESC \cite{angulo_hahn}.

One way to suppress the cosmic variance in these simulations was introduced by \cite{Angulo16}, and is commonly referred to as "Fixing and Pairing" (hereafter F\&P). The procedure consists of two steps; the first is to generate initial conditions with amplitudes fixed to their ensemble-average values, so that the matter power spectrum of one realization will be exactly equal to the the ensemble mean. Secondly, one combines two such simulations with the same initial seed, but shifting the phases of one of them by $\pi$, or equivalently, generating it with opposite initial linear overdensities; the deviations of their two-point functions with respect to the ensemble average will be roughly the same, only with opposite sign, so that averaging the two is expected to make this statistic converge to the ensemble average much faster than it would by simply averaging two realizations with random seeds. A wide range of statistics derived from the matter density field in F\&P simulations have been investigated, finding that they are unbiased and that some feature a very significant reduction in their variances \cite{Angulo16, Villaescusa_Navarro_2018, Klypin_2020}.

Reducing the variance in matter statistics is a great advancement, but interpreting observations requires models for biased tracers, and thus one must test whether their statistics also remain unbiased and quantify the variance suppression. This has been done by \cite{Villaescusa_Navarro_2018} using halos in N-Body simulations, or directly hydrodynamic simulations, seeing that these statistics are unbiased and display some variance suppression, albeit smaller than for matter. Alternatively, one can model these tracers by combining several properties of the dark-matter density field with a set of bias parameters \cite{Desjacques:2016bnm}. It has not been checked whether this can be done using the results of F\&P simulations without introducing biases, nor how much the statistical improvements are retained in the final spectrum. This is particularly relevant for current developments of theoretical models of galaxy clustering based on combining numerical simulations and the bias expansion, as F\&P is a valuable tool in making them more precise \cite{kokron2021cosmology, zennaro2021bacco, hadzhiyska2021hefty, Pellejero_2022, Pellejero_2022B}.

Theoretical models are never a perfect description of reality, and one must have good estimates of the error being made so that it can be included in the data analyses, avoiding distorted interpretations of the measurements \cite{baldauf2016, Sprenger2019}. In the case of models based on numerical simulations, these errors may come from cosmology rescaling methods \cite{Angulo_2010}, emulators of statistical quantities, and from the undersampling of large-scale modes in the simulation itself. The latter can in principle be reduced using F\&P, which helps make the model more precise, but hinders our capacity of predicting its error \cite{Zhang:2021hrq} -- one knows only that the variance is smaller, but not its value.

In this work, we set out to understand the statistical properties of models of biased tracers built from F\&P simulations, answering several of the questions raised in the preceding paragraphs. We will assess whether F\&P introduces biases to the basis spectra entering the bias expansion, and quantify the variance suppression for each of them. Furthermore, we will provide qualitative explanations for their different behaviours and quantitative predictions capable of describing their variances. We also propose and test a method to further reduce the variances in certain spectra that are unaffected by F\&P, and finally apply the developed tools to understand what the minimum simulation size is such that the error in a theoretical model built from it satisfies EUCLID requirements.

The text of this manuscript is organized as follows. Section \ref{section:fixed_and_paired} is devoted to establishing our notation, and to reviewing the fixing and pairing methods. Section \ref{section:bias_expansion} gives a very brief account of the bias expansion, and explains the way in which it has been applied in this work. In section \ref{section:numerical_results} we describe the set of simulations utilized throughout this work and employ them to demonstrate that the basis spectra entering the bias expansion are unbiased in F\&P simulations, and to quantify the reduction in their variances. In Section \ref{section:th_exploration} we use Lagrangian Perturbation-Theory (hereafter LPT) to understand why certain spectra feature noise reduction in F\&P simulations, while others do not, and to obtain, for the first time, simple analytical expressions for the variances of these spectra in F\&P simulations. In section \ref{section:ph_field} we show how to further reduce the variance in the basis spectra. Finally, section \ref{section:euclid_forecast} is devoted to applying these results to understand the simulation sizes required to build a model precise enough for the analysis of Euclid data, and in section \ref{section:conclusion} we conclude and present final remarks.

\section{Fixed and Paired Fields}
\label{section:fixed_and_paired}
In this section we give a brief overview to contextualize F\&P fields, and to settle on the notation to be used throughout this work.

Let us say $\rho(\bm{x})$ is the density of dark matter, then we can define the matter overdensity field as $\delta(\bm{x}) = \frac{\rho(\bm{x})}{\bar{\rho}} - 1$. It is customary to work with its Fourier transform, defined by
\begin{equation}
    \delta(\bm{k}) = \int d^3\bm{x} \delta(\bm{x})e^{-i\bm{k}\cdot\bm{x}}.
\end{equation}
This field will be complex, and therefore can be decomposed into an amplitude and a phase, $\delta(\bm{k}) = |\delta(\bm{k})|e^{i\theta_{\bm{k}}}$; due to $\delta(\bm{x})$ being real, these fields must satistfy $\delta^*(\bm{k})=\delta(-\bm{k})$ and consequently $\theta_{-\bm{k}} = -\theta_{\bm{k}}$. One can in general define the power spectrum of this field through the expression
\begin{equation}
    \langle \delta(\bm{k})\delta(\bm{k}') \rangle = (2\pi)^3 \delta_D(\bm{k}-\bm{k}') P(k),  
\end{equation}
in which we express the power spectrum directly as depending only on the modulus of the wave vector due to the assumption of isotropy in the density field. In the case of a numerical simulation with finite volume $L^3$, the power spectrum can be defined as
\begin{equation}
    P(\bm{k}) = \frac{1}{L^3} \delta(\bm{k}) \delta(-\bm{k}),
\end{equation}
and isotropy is only valid in the limit of $L^3\rightarrow \infty$, so in order to obtain $P$ as a function of the modulus of the wave-vector, one must average over a spherical shell of radius $k_i$ and width $\Delta k$,
\begin{equation}
    P(k_i) = \frac{1}{V_s(k_i)} \int_{\bm{k}_i} \frac{d^3\bm{k}}{(2\pi)^3} P(\bm{k}),
\end{equation}
in which $V_s(k_i) = 4\pi k_i^2 \Delta k = 4\pi k_i^3 \Delta(\ln k)$, depending on whether one uses linear or logarithmic binning of the wave modes, and $\int_{\bm{k}_i}$ is an integral over $V_s(k_i)$, the Fourier shell of radius $k_i$ and width $\Delta k$.

Observations of the cosmic microwave background \cite{planck2018} have shown that perturbations in the early universe were, to a very good approximation, Gaussianly distributed. This can be expressed equivalently by saying that the absolute values of the Fourier modes follow a Rayleigh distribution, and the phases are uniformly distributed in the interval $[0,2\pi]$, resulting in a combined distribution given by
\begin{equation}
\mathcal{P}(|\delta(\bm{k})|, \theta_{\bm{k}}) = \frac{|\delta|}{\pi L^3 P} e^{-|\delta|^2/L^3 P}.
\end{equation}
Therefore, the standard procedure for generating initial conditions for cosmological simulations is to sample this distribution for amplitudes and phases, using the linear power spectrum at high redshift in place of $P$. To generate fixed simulations, in contrast, one substitutes this probability density function by a Dirac delta centered at $\sqrt{L^3P(\bm{k})}$, that is,
\begin{equation}
    \mathcal{P}^{f}(|\delta|(\bm{k})|,\theta_{\bm{k}}) = \frac{1}{2\pi}\delta_D\left(|\delta| - \sqrt{L^3P}\right),
\end{equation}
which amounts to fixing all the amplitudes to the value $|\delta(\bm{k})| = \sqrt{L^3P(\bm{k})}$ while allowing the phases to be uniformly sampled between $0$ and $2\pi$. From this initial density field, $\delta_1(\bm{k})=|\delta(\bm{k})|e^{i\theta_{\bm{k}}}$, one can easily generate its corresponding pair simply by shifting the phases of $\pi$, $\delta_2(\bm{k})=|\delta(\bm{k})|e^{i(\theta_{\bm{k}}+\pi)}$. Fixing clearly reduces the level of randomness in the field while destroying its Gaussianity. Nevertheless, works such as \cite{Angulo16, Villaescusa_Navarro_2018, Klypin_2020} have done extensive studies of many statistical quantities derived from F\&P simulations, finding that two and three-point clustering statistics, as well as the mass function and probability density function are unbiased; biases appear in the variance and covariance of two-point functions -- as intended, since the method has been designed to suppress these quantities.

\begin{figure}[tbp]
\centering 
\includegraphics[width=\textwidth]{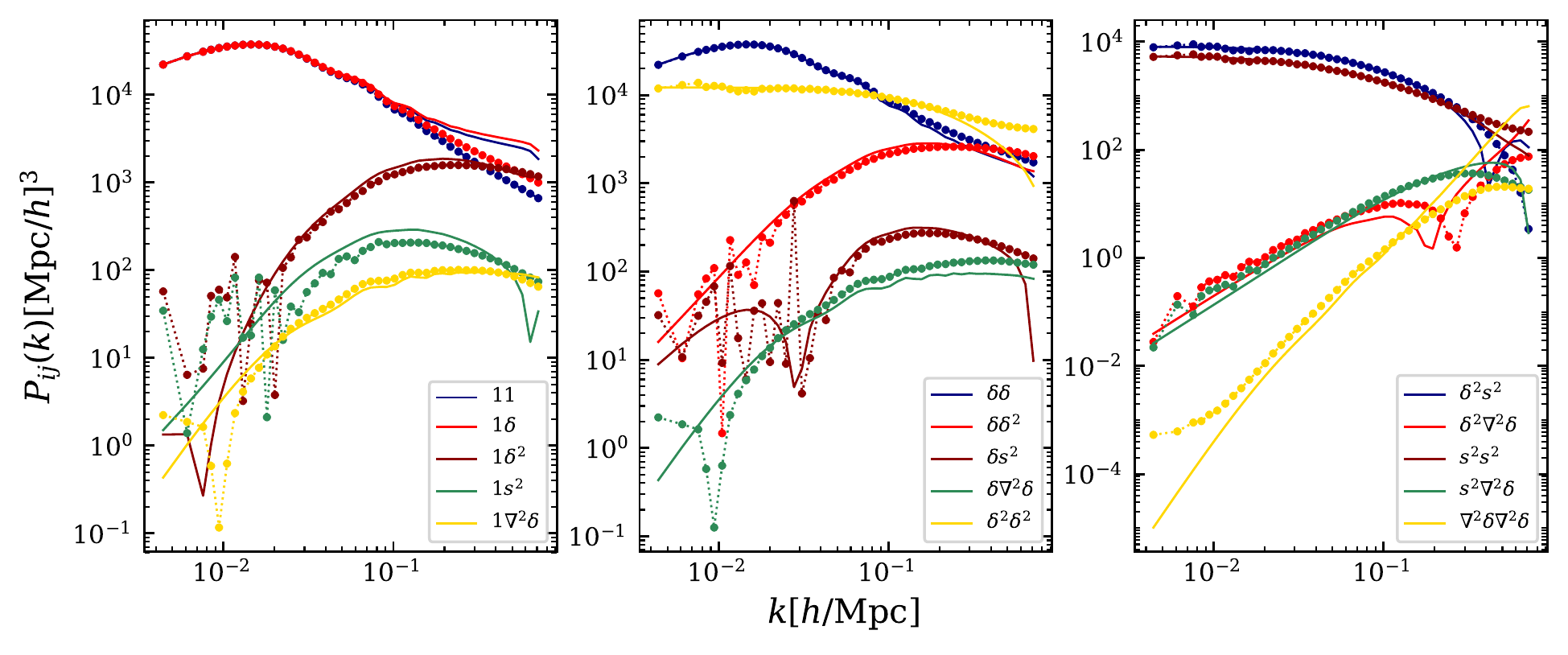}
\caption{Basis spectra entering a second-order Lagrangian bias expansion, advected to Eulerian space. Solid lines represent calculations using first-order LPT, and filled circles represent the mean of these spectra measured from a set of 200 COLA simulations.}
\label{fig:basis_specs}
\end{figure}

\section{Bias Expansion}
\label{section:bias_expansion}

Biased tracers of the large-scale structure can be modeled in many ways, ranging from detailed and expensive methods such as finding halos or galaxies in simulations, or doing subhalo abundance matching to populate dark-matter simulations with galaxies, to simpler and less detailed methods such as a perturbative bias expansion. In this work we will model galaxies by applying a second-order Lagrangian bias expansion with an additional term to account for nonlocal effects of galaxy formation,
\begin{equation}
    1+\delta_g(\bm{q}) = 1 + b^L_1 \delta(\bm{q}) + b_2^L \delta^2(\bm{q}) + b_{s^2}^L s^2(\bm{q}) + b_{\nabla^2\delta}^L \nabla_{\bm{q}}^2\delta(\bm{q}) + \cdots,
    \label{eq:bias_exp}
\end{equation}
in which $s^2(\bm{q}) = s_{ij}s_{ij}$ and $s_{ij}$ is the traceless part of the tidal tensor. This has the advantage of being agnostic to the galaxy formation processes, and also very flexible, allowing one to fit virtually any galaxy population merely by changing the values of the bias parameters \cite{Desjacques:2016bnm}. 

Using equation (\ref{eq:bias_exp}) one can obtain an approximation for the galaxy density field in Lagrangian space; however, to compare it to observations one must have it in Eulerian space. To perform this transformation, we must move particles from their Lagrangian to Eulerian positions,
\begin{equation}
    \begin{split}
        1 + \delta_g(\bm{x}) & = \int d^3\bm{q} \delta^D(\bm{x}-\bm{q}-\bm{\psi}(\bm{q})) (1 + \delta_g(\bm{q}) ) \\
        & = \sum_{F(\bm{q}) \in \left[1, \delta, \delta^2, s^2, \nabla^2\delta \right]} b_{F} \int d^3\bm{q} \delta^D(\bm{x}-\bm{q}-\bm{\psi}(\bm{q}))F(\bm{q}) \\
        & = \sum_{F(\bm{q}) \in \left[1, \delta, \delta^2, s^2, \nabla^2\delta \right]} b_{F} \delta_F(\bm{x}), 
    \end{split}
    \label{eq:bias_exp_advect}
\end{equation}
in which we have used the linearity of the bias expansion to reorganize the expression for the galaxy overdensity in Eulerian space as a linear combination of advected Lagrangian operators. From this, one can then obtain an expression for the galaxy power spectrum expressed in terms of the bias coefficients and the auto and cross-spectra of the advected dark-matter density fields,
\begin{equation}
    P_{gg}(k) = \sum_{i,j \in \left[1, \delta, \delta^2, s^2, \nabla^2\delta \right]} b_ib_j P_{ij}(k).
    \label{eq:spec_bias_exp}
\end{equation}
Notice that in this expression $b_1$ must be equal to $1$, and is merely a bookkeeping parameter; in the next sections we will refer to the bias parameter denoted here by $b_{\delta}$ as $b_1^L$, that is, the linear Lagrangian bias. Recent work \cite{Modi_2020, kokron2021cosmology, zennaro2021bacco, hadzhiyska2021hefty, zennaro2021priors, Kokron_21, Pellejero_2022, Pellejero_2022B} has shown that one can use this modeling and perform the advection using the nonlinear displacement field predicted by simulations to obtain accurate predictions for the power spectra of biased tracers at highly nonlinear scales. We call this general procedure "hybrid Lagrangian bias expansion".

\section{Numerical Results}
\label{section:numerical_results}

To perform our numerical analyses we have run an ensemble of 200 L-PICOLA simulations, which were used for comparing Gaussian and F\&P spectra, and for understanding up to which scales our theoretical predictions are valid; this set of simulations was run with the parameters resumed in table (\ref{COLA_pars}), for Gaussian and fixed initial conditions, and for phase offsets $\phi=0,\pi$. The cosmological parameters correspond to those used in the Millennium simulation \cite{millenium}; the parameters $n_{steps}$, $N_{sample}$ and $N_{part}$ control, respectively, the time, force, and mass resolutions of the simulation. The original L-PICOLA code developed by \cite{howlett2015lpicola} did not include an option for fixing or pairing, and therefore, minor modifications had to be done to it.

In order to measure the basis spectra entering equation (\ref{eq:spec_bias_exp}) we took the following procedure:
\begin{enumerate}
    \item Generate initial conditions with the same seed and phase offset as the simulation to be analysed;
    \item Use this generated field to compute $\delta(\bm{q}), \delta^2(\bm{q}), s^2(\bm{q}) \text{ and } \nabla_q^2\delta(\bm{q})$ in Lagrangian space;
    \item Compare the initial positions of the particles to their positions at the desired redshift, thus obtaining their displacement $\bm{\psi}_i = \bm{x}_i(z_{final}) - \bm{x}_i(z_{init})$;
    \item Advect the quantities from Lagrangian to Eulerian space using the procedure outlined in equation \ref{eq:advect}
    \item Compute their auto and cross-spectra using the fields obtained in Eulerian space. For the paired spectra we then average over the two simulations run with the same seed, but different phase offsets.
\end{enumerate}
The mean of these spectra is shown in figure (\ref{fig:basis_specs}), along with theoretical predictions obtained using 2LPT.

\begin{table}[t!]
\centering
\begin{tabular}{|c|c|}
\hline
$\Omega_m$ & $0.25$ \\
$\Omega_b$ & $0.045$\\
$\Omega_\Lambda$ & $0.75$\\
$h$ & $0.73$\\
$\sigma_8$ & 0.90\\
$n_s$ & $0.96$\\
$L$ & $1.5$ Gpc$/h$\\
$z_i$ & $9$\\
$n_{steps}$ & $10$\\
$N_{sample}$ & $400$\\
$N_{mesh}$ & $800$\\
\hline
\end{tabular}
\label{COLA_pars}
\caption{ Parameters used to run the ensemble of L-PICOLA simulations used in this work.}
\end{table}

\subsection{Bias and Variance Reduction}

In this section, we use the previously mentioned COLA simulations to numerically assess by how much fixing and/or pairing reduce the variance in the basis spectra, and whether these spectra are unbiased compared to those obtained using Gaussian initial conditions.

\begin{figure}[t!]
    \centering
    \includegraphics[width=\textwidth]{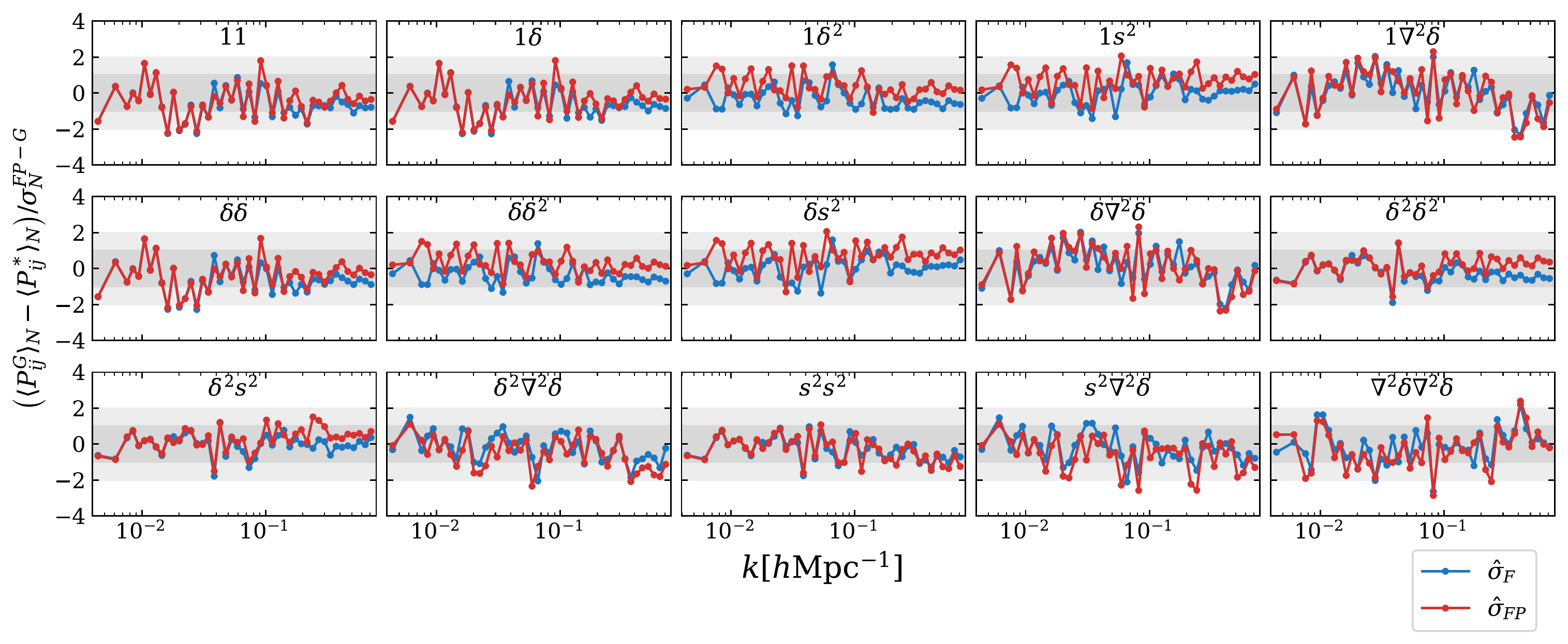}
    \caption{Differences between means of $N=20$ fixed and/or paired spectra to Gaussian ones, divided by their standard deviation, as estimated from an ensemble of 200 COLA simulations using equation (\ref{eq:sigma_N}). These plots indicate clearly that there are no significant biases introduced by F\&P, since the measured differences generally remain inside the 1 or 2-$\sigma$ region (indicated by grey shaded regions).}
    \label{fig:bias_cola}
\end{figure}

Using the simulation results, we estimate the bias of F\&P simulations with respect to Gaussian ones by using the expression
\begin{equation}
    B^{FP}_N = \frac{\langle P^{FP} \rangle_N - \langle P^{G} \rangle_N }{\sigma_N^{FP-G}},
\end{equation}
in which $\langle P^{FP} \rangle_N$ is the average of $N$ F\&P simulations, and analogously for $\langle P^G \rangle_N$; $\sigma_N^{FP-G}$ is the expected error in their difference, defined by the expression
\begin{equation}
    \sigma^{FP-G}_N = \sqrt{\left[\sigma_N^{fp}\right]^2 + \left[\sigma_N^{G}\right]^2},
\end{equation}
and the standard deviations $\sigma_N^{FP}$ and $\sigma_N^G$ are the estimated errors on the average of $N$ spectra, defined by
\begin{equation}
    \left[\sigma_N^{FP}\right]^2 = \frac{1}{n-1} \sum_{i=1}^n \left( \langle P^{FP} \rangle_{N_i} - \langle P^{FP} \rangle\right)^2,
    \label{eq:sigma_N}
\end{equation}
$n$ being the number of independent sets of $N$ spectra available. In figure (\ref{fig:bias_cola}) we show the results for the computation of these biases from 10 sets of 20 simulations; it is not hard to see that, for all of the 15 basis spectra, no significant bias can be detected, as most of the points are located inside the $1$ or $2-\sigma$ region, represented by gray-shaded areas in the figure. This result was expected from analytical considerations made in \cite{Angulo16}, in which the authors show that the procedure of fixing the amplitudes of the initial Fourier modes is expected to change the power spectra only at a measure-zero set of their domain.

To quantify the statistical improvement derived from fixing we directly take the ratio between the variances of Gaussian and fixed spectra, $\frac{\sigma_{G}}{\sigma_{f}}$. Quantifying the statistical improvement coming from pairing is slightly more delicate. Paired spectra are obtained from the average of spectra from two simulations with same seed, but opposite initial conditions; if these spectra were uncorrelated, then the variance of the average would be divided by $2$. Of course this would not be much of an improvement since we could just have run an additional simulation, with a different seed, to get the same effect; the interesting case is when, due to having opposed ICs, the spectra from these simulations are anticorrelated and their fluctuations around the mean will cancel out making the averaged spectra converge to the ensemble mean much faster. Therefore, we choose to multiply the variance of paired simulations by $2$ before comparing them to the other ones, to avoid seeing the effects of simply averaging over two simulations. A side effect to this is that, for some spectra, the measurements from the two simulations are highly correlated so that the second simulation adds almost no information, giving only a small reduction in variance from taking their average; multiplying it by $2$, can then make it artificially larger than the variance of the non-paired spectra, which does not mean that pairing worsened the variance, but only that the second simulation added almost no new information.

\begin{figure}[tbp]
    \centering
    \includegraphics[width=\textwidth]{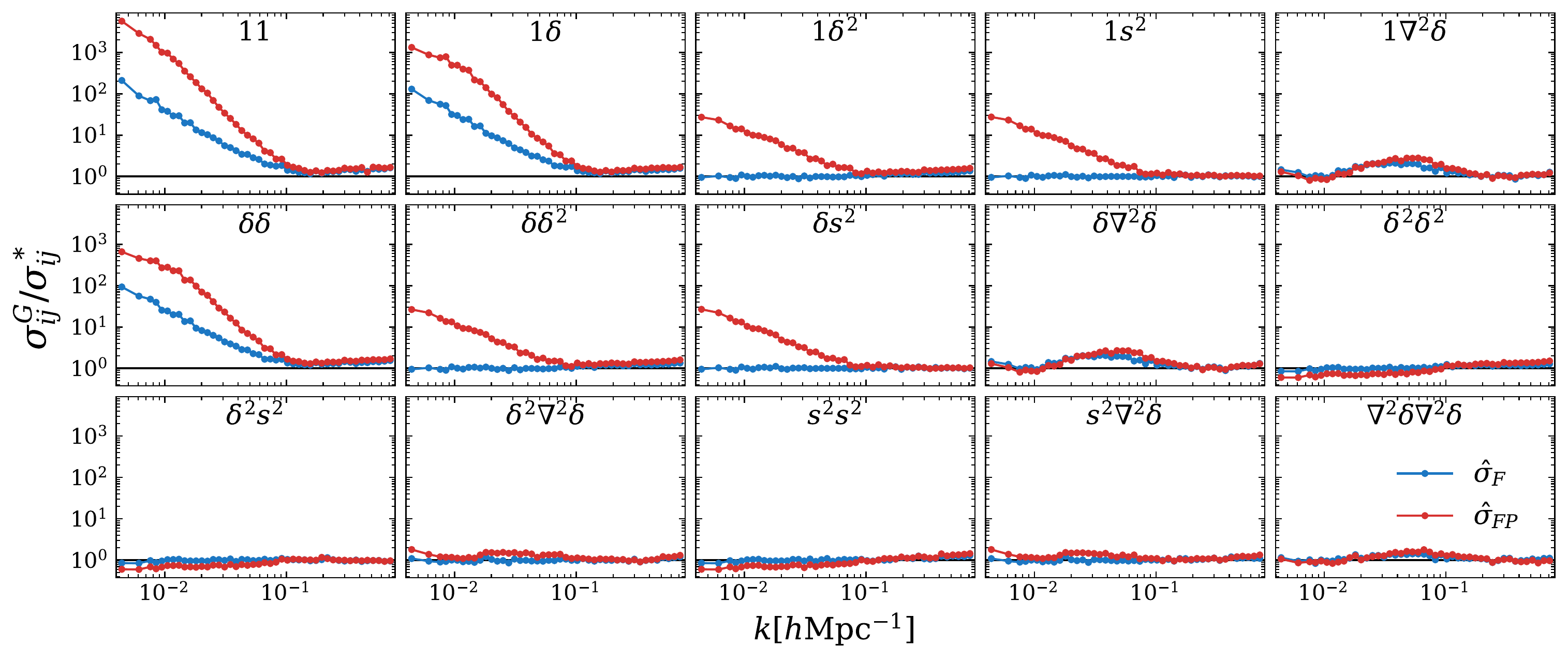}
    \caption{Ratio of the standard-deviation in standard spectra to the same quantity for fixed and/or paired ones. One can see how some of these spectra profit from a large reduction in variance from both fixing and pairing. Others profit only from pairing, while a third group remains unchanged by either procedure. Also worth remarking is the behaviour of spectra containing $\nabla^2 \delta$ which is somewhat undefined. }
    \label{fig:variance_ratios}
\end{figure}

Figure (\ref{fig:variance_ratios}) shows the analysis assessing the variance reductions from F\&P. It is useful to divide these spectra into three categories to better understand the reduction in variance: 

\begin{itemize}
    \item Spectra involving two linear fields, $P_{11}, P_{1\delta}, P_{\delta\delta}$. For scales $k\sim 10^{-2}[h/$Mpc$]$ fixing gives a large reduction in $\sigma$ of a factor of around $10^2$ at the largest scales, which then reduces until no improvement is seen at scales $k\approx 0.15[h/$Mpc$]$; a small improvement is seen for $k \gtrsim 0.15[h/$Mpc$]$, but is not very relevant both because of its much smaller magnitude, and because the error in the spectra at these scales is already quite small. At the largest scales pairing further reduces $\sigma$ by a factor of $10$ and this reduction diminishes as we go to larger $k$, until no improvement is seen beyond $k\approx 0.15[h/$Mpc$]$.
    
    \item Spectra involving one linear and one squared field, $P_{1\delta^2}, P_{1s^2}, P_{\delta\delta^2}, P_{\delta s^2}$. Fixing gives no benefit for these spectra at scales $k\lesssim 0.15$, and for scales beyond this threshold we see only a very small improvement for $P_{1\delta^2}$ and $P_{\delta\delta^2}$. Pairing gives considerable reduction in $\sigma$ of a factor of $10$ at scales $k\sim 10^{-2}[h/$Mpc$]$, which then falls down until no improvement is seen for $k\gtrsim 0.15[h/$Mpc$]$.
    
    \item Spectra involving two squared fields $P_{\delta^2\delta^2}, P_{\delta^2s^2}, P_{s^2s^2}$. For these spectra we see no improvement from fixing or pairing.
    
\end{itemize}
A comment is in order regarding the spectra involving the Laplacian field $P_{1\nabla^2\delta}$, $P_{\delta\nabla^2\delta}$, $P_{\delta^2\nabla^2\delta}$, $P_{s^2\nabla^2\delta}$, $P_{\nabla^2\delta\nabla^2\delta}$. For these spectra we see a mild improvement from fixing and pairing at scales $10^{-2} \lesssim k/[h$Mpc$^{-1}]\lesssim 0.1$, but we wish not to include it in any of the categories above both because they present an irregular behaviour, with no clear tendency as a function of scale, and because in any way it is not relevant to predict these variances because the corresponding spectra are extremely suppressed with respect to the others. In the next section we will provide simple explanations to these three classes of behaviours, and compute their variances analytically.

\section{Analytic Exploration}
\label{section:th_exploration}

In order to understand the results exposed above, we will use the LPT formalism \cite{Bernardeau:2001qr, Matsubara_2008, rampf2012, Carlson_2012, Sugiyama_2014, rampf2021} to compute approximate expressions for the power spectra that go into equation (\ref{eq:spec_bias_exp}) and their variances; this will allow us to see the effects of fixing or pairing on the structure of these quantities. Let us call $F(\bm{q})$ a field in Lagrangian space such that 
\begin{equation}
    F(\bm{q}) \in [1, \delta, \delta^2, s^2, \nabla^2\delta].
\end{equation}
We are interested in these quantities in Eulerian space, and therefore we must advect them using the displacement field, as previously described in equation (\ref{eq:bias_exp_advect}),
\begin{equation}
    \delta_F(\bm{x}) = \int d^3\bm{q} F(\bm{q}) \delta_D( \bm{x} - \bm{q} - \bm{\psi}(\bm{q})).
\end{equation}
Taking the Fourier transform of both sides and expanding the displacement to first order in perturbation theory, we obtain the following results.
\begin{equation}
\begin{split}
    \delta_F(\bm{k}) \approx & \int d^3\bm{q} e^{-i\bm{k}\cdot\bm{q}} F(\bm{q}) e^{-i\bm{k} \cdot\bm{\psi}^{(1)}(\bm{q}) } \\
     \approx & \int d^3\bm{q} e^{-i\bm{k}\cdot\bm{q}} F(\bm{q}) \left[1 - i\bm{k}\cdot\bm{\psi}^{(1)} - \frac{(\bm{k}\cdot\bm{\psi}^{(1)})^2}{2} \right]\\
     \label{eq:advect}
\end{split}
\end{equation}
in which going to the second line we Taylor expanded the exponential up to second order; one can then plug in the Z'eldovich approximation $\psi(\bm{k})=i\frac{\bm{k}}{k^2}\delta_L(\bm{k})$ and the different fields into this expression and perform the relevant calculations to obtain them in Eulerian space. We notice that, although the results to be presented in sections 5.1 - 5.4 are obtained using equation (\ref{eq:advect}) to advect the fields, our qualitative conclusions are completely general and rely only on the structure of perturbative calculations. 

With the expressions of $\delta_F(\bm{k})$ at hand, we can obtain the auto and cross power spectra of these fields through
\begin{equation}
    \widehat{P}_{\mu\nu}(\bm{k}) = \frac{1}{V} \text{Re} \left[\delta_\mu(\bm{k})\delta_\nu(-\bm{k}) \right],
\end{equation}
in which $\mu,\nu\in\left[1,\delta,\delta^2,s^2,\nabla^2\delta\right]$ and $V=L^3$. Notice that to make this calculation fully consistent up to 1-loop, we should have retained several terms in equation (\ref{eq:advect}) which we have discarded; this is justified simply because our purpose is not to obtain highly precise expressions for these variances but to provide insight into the mechanism by which these methods operate, and to validate our approach showing we can make reasonable quantitative predictions. We attract the attention of the reader to the fact that we are not taking the ensemble average of this combination of fields. This is done with the purpose of retaining terms which would be zero in the ensemble average but are present for simulations due to their finite size; these terms will be very important in our explanations of the variance reduction in different spectra.

\subsection{Results for $P_{11}, P_{1\delta}, P_{\delta\delta}$}

Using equations (\ref{eq:delta_1}) and (\ref{eq:delta_delta}) one can easily obtain the expression for $P_{11}$ in the case of fixed initial Fourier amplitudes \footnote{In the following sections, whenever an integral of the kind $\int \frac{d^3\bm{k}}{(2\pi)^3}$ appears, we will simplify the notation, substituting it by $\int \frac{d^3\bm{k}}{(2\pi)^3} \longrightarrow \int_{\bm{k}}$.}
\begin{equation}
    \begin{split}
        P^{F}_{11}(\bm{k}) \approx &  P^L_{\bm{k}} + V^{1/2}\int_{\bm{q}_1}\sqrt{P^L_{\bm{k}}P^L_{\bm{q}_1}P^L_{\bm{q}_1-\bm{k}}}\cos\left[\theta_{\bm{k}} - \theta_{\bm{q}_1} - \theta_{\bm{q}_1-\bm{k}}\right]F_{ZA}(\bm{q}_1,\bm{k}-\bm{q}_1, \bm{k}) \\
        & + \frac{V}{4}\int_{\bm{q}_1}\int_{\bm{q}_2} \sqrt{P^L_{\bm{q}_1}P^L_{\bm{k}-\bm{q}_1}P^L_{\bm{q}_2}P^L_{\bm{q}_2-\bm{k}}} \cos\left[ \theta_{\bm{q}_1} + \theta_{\bm{k}-\bm{q}_1} - \theta_{\bm{q}_2} - \theta_{\bm{k}-\bm{q}_2} \right] \\
        & \times F_{ZA}(\bm{q}_1,\bm{k}-\bm{q}_1, \bm{k})F_{ZA}(\bm{q}_2,\bm{k}-\bm{q}_2, \bm{k}).
    \end{split}
    \label{eq:p_11}
\end{equation}
The leading-order contribution to this spectrum is the linear power spectrum, $P^L_{\bm{k}}$, and for the fixed case this has no variance whatsoever; indeed, when we combined the two fixed fields, their phases canceled completely, and since their amplitudes are fixed to the square root of the linear power spectrum, we obtain it completely free of noise. It is not hard to see that, for all spectra which have the linear power spectrum as their leading order term, fixing will greatly suppress their variance, by cancelling its leading order contribution. This is precisely the case of these spectra,
\begin{equation}
    P_{11}, P_{1\delta}, P_{\delta\delta} \supset P^L,
\end{equation}
explaining why fixing provides a large reduction to their variances.

The next-to-leading-order contribution to $P_{11}$ is given by a term which came from the combination of three factors of $\delta^L$, and we will thus denote it schematically by $(\delta\delta\delta)$. Taking a closer look at this term we can see that, for a simulation which had all of its phases displaced by $\pi$, one would obtain
\begin{equation}
\begin{split}
    (\delta\delta\delta)_\pi \sim & \int_{\bm{q}_1}\sqrt{\cdots}\underbrace{\cos\left[\theta_{\bm{k}} - \theta_{\bm{q}_1} - \theta_{\bm{q}_1-\bm{k}} - \pi\right]}_{-\cos\left[\theta_{\bm{k}} - \theta_{\bm{q}_1} - \theta_{\bm{q}_1-\bm{k}}\right]}F_{ZA}(\bm{q}_1,\bm{k}-\bm{q}_1, \bm{k}) \\
    = & -(\delta\delta\delta).
\end{split}
\end{equation}
The pairing procedure consists precisely of creating two simulations with the same seed -- but one of them with all the phases displaced by $\pi$ -- and then average their spectra. Through the arguments above one can clearly see that this would cancel the $(\delta\delta\delta)$ -- along with all other combinations of an odd number of factors of the density field -- from the spectrum. Although the explanation we gave was focused on the case of $P_{11}$, it is valid for the other two spectra, as all of these contain a term composed of three linear fields, which has a completely analogous structure to the one described above. Schematically, one can say that
\begin{equation}
    P_{11}, P_{1\delta}, P_{\delta\delta} \supset (\delta\delta\delta),
\end{equation}
and therefore, all of these spectra have their variances further reduced by pairing. 

The only term at this order of the calculation which continues to contribute to the variance is
\begin{equation}
    \begin{split}
        \label{eq:left_sig11}
        \frac{V}{4}\int_{\bm{q}_1}\int_{\bm{q}_2} \sqrt{P^L_{\bm{q}_1}P^L_{\bm{k}-\bm{q}_1}P^L_{\bm{q}_2}P^L_{\bm{q}_2-\bm{k}}} \cos\left[ \theta_{\bm{q}_1} + \theta_{\bm{k}-\bm{q}_1} - \theta_{\bm{q}_2} - \theta_{\bm{k}-\bm{q}_2} \right] \\
        \times F_{ZA}(\bm{q}_1,\bm{k}-\bm{q}_1, \bm{k})F_{ZA}(\bm{q}_2,\bm{k}-\bm{q}_2, \bm{k}).
    \end{split}
\end{equation}
A very interesting question is then raised, regarding how can one reduce the variance associated with terms such as this one, and this expression allows us to immediately see that searching for other phase-translations in the same spirit as the pairing procedure will not help. This is because the cosine in the above expression has two phases with a plus sine, and two with a minus, so that any phase translation will be canceled and will not affect this term whatsoever.

\subsection{Results for $P_{1\delta^2}, P_{1s^2}, P_{\delta\delta^2}, P_{\delta s^2}$}

From Appendix \ref{section:advected_fields} we can take the expression for the fields $\delta_1$ and $\delta_{\delta^2}$, and compute $P_{1\delta^2}$ up to fourth order in $\delta_L$ in the case of fixed initial Fourier amplitudes; this will be given by
\begin{equation}
    \begin{split}
        P^{F}_{1\delta^2}(\bm{k}) \approx & V^{1/2} \int_{\bm{q}_1}\sqrt{P^L_{\bm{k}} P^L_{-\bm{q}_1}P^L_{\bm{q}_1-\bm{k}}} \cos\left[\theta_{\bm{k}} - \theta_{\bm{q}_1} - \theta_{\bm{k}-\bm{q}_1} \right] \\
        & + V\int_{\bm{q}_1}\int_{\bm{q}_2} \frac{\bm{k}\cdot(\bm{k}-\bm{q}_{12})}{|\bm{k}-\bm{q}_{12}|^2} \sqrt{P^L_{\bm{k}}P^L_{\bm{q}_1}P^L_{\bm{q}_2}P^L_{\bm{k}-\bm{q}_{12}}}\cos\left[ \theta_{\bm{k}} - \theta_{\bm{q}_1} - \theta_{\bm{q}_2} - \theta_{\bm{k}-\bm{q}_{12}}\right]\\
        & + \frac{V}{2}\int_{\bm{q}_1}\int_{\bm{q}_2}\mathcal{K}_1(\bm{q}_1,\bm{k}-\bm{q}_1) \sqrt{P^L_{\bm{q}_1}P^L_{\bm{k}-\bm{q}_1}P^L_{\bm{q}_2}P^L_{\bm{k}-\bm{q}_2}}\cos\left[ \theta_{\bm{q}_1} + \theta_{\bm{k}-\bm{q}_1} - \theta_{\bm{q}_2} - \theta_{\bm{k}-\bm{q}_2} \right]
    \end{split}
    \label{eq:linear-squared}
\end{equation}
This makes it clear that the leading order contribution to this spectrum is not coming from the linear power-spectrum; indeed this supports our observations from numerical simulations, which show that fixing has no effect in reducing the variance of these spectra. The intense mode-mixing coming from the integrals in equation (\ref{eq:linear-squared}) destroys the regularity introduced by fixing the amplitudes, leaving the statistical noise unchanged. However, we see that the leading-order term for these spectra is in fact precisely of the shape $(\delta\delta\delta)$, allowing us to affirm
\begin{equation}
    P_{1\delta^2}, P_{\delta \delta^2}, P_{1 s^2}, P_{\delta s^2} \supset (\delta\delta\delta), 
\end{equation}
and thus the argument developed in the preceding subsection is capable of explaining the substantial variance reduction to these spectra coming from pairing.

\subsection{Results for $P_{\delta^2\delta^2}, P_{s^2s^2}$, $P_{\delta^2s^2}$}

Analogously to what has been done before, one can take expressions for $\delta_{\delta^2}$ from appendix \ref{section:advected_fields} and compute $P_{\delta^2\delta^2}$ for the case of fixed initial Fourier amplitudes; this gives us
\begin{equation}
    P^{F\&P}_{\delta^2\delta^2} \approx \frac{1}{V_f} \int_{\bm{q}_1} \int_{\bm{q}_2} \sqrt{P^L_{\bm{q}_1}P^L_{\bm{k}-\bm{q}_1}P^L_{\bm{q}_2}P^L_{\bm{k}-\bm{q}_2}} \cos\left[ \theta_{\bm{q}_1} + \theta_{\bm{k}-\bm{q}_1} -\theta_{\bm{q}_2} - \theta_{\bm{k}-\bm{q}_2} \right].
\end{equation}
The leading-order contribution to this spectrum is an integral which will mix the initial Fourier modes in a highly nontrivial way, and will thus be unaffected by fixing. Pairing will not affect this term either since the first odd combination of factors of the density field will only appear at fifth order, and will therefore be negligible in comparison to the contribution from the fourth order term. It is interesting to notice that this term has many similarities in form with equation (\ref{eq:left_sig11}) and with one of the fourth-order contributions to equation (\ref{eq:linear-squared}). If one was to encounter a method analogous to F\&P to mitigate the variance associated with this kind of term, there would be a massive reduction in the error of the final galaxy clustering model due to the simultaneous reduction in all the relevant basis spectra. 

\subsection{Discussion on $P_{1\nabla^2\delta}, P_{\delta\nabla^2\delta}, P_{\delta^2\nabla^2\delta}, P_{s^2\nabla^2\delta}, P_{\nabla^2\delta\nabla^2\delta}$}

Analytically, the behaviour of these terms would be very simple to derive, since in Fourier space one can write $\nabla^2\delta \rightarrow -k^2\delta$. It is quite clear that, although multiplying $\delta$ by $k^2$ could change the kernels connecting the linear fields, it would not change the general structure of these terms, which should then behave as if the laplacian was a linear field; as stated earlier, this is not what we see. It is not yet clear whether these effects are a product of a lack of numerical resolution to which the laplacian would be exceedingly sensible, or if it would be necessary to improve our analytical modeling to reproduce this behaviour.

\subsection{Quantitative Predictions}

\begin{figure}[tbp]
\centering 
\includegraphics[width=\textwidth]{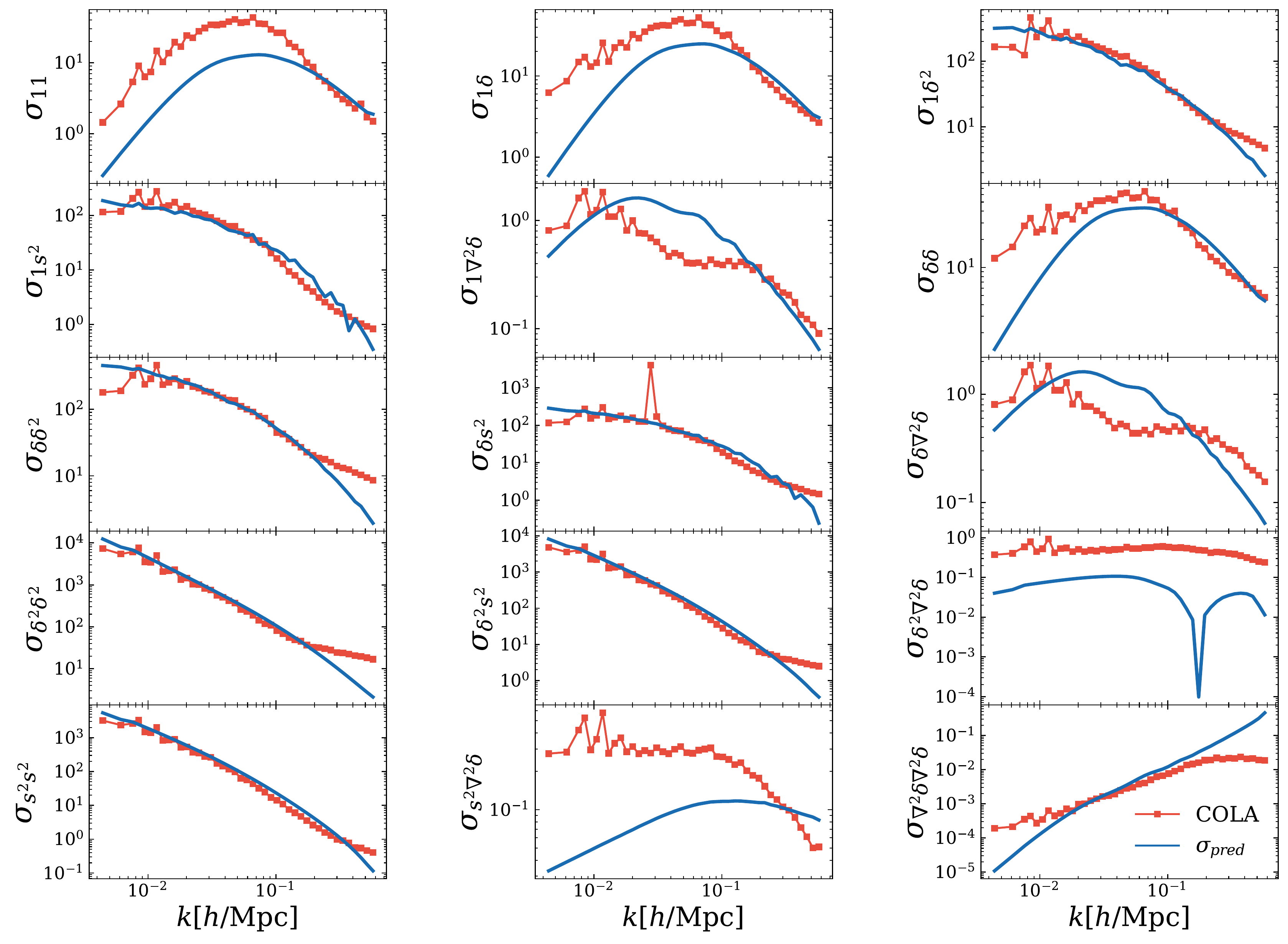}
\caption{Theoretical predictions for the variances compared to measurements from an ensemble of 200 COLA simulations. }
\label{fig:cola_var}
\end{figure}

Throughout this section we have presented a theoretical framework for analytically computing power spectra while retaining in these expressions the statistical properties characteristic of a simulation, and have employed it to obtain a qualitative understanding on the behaviour of the variances for auto and cross-spectra in F\&P simulations. This same framework can also be used to compute expressions for the variances of these spectra in F\&P simulations, an thus obtain quantitative predictions for them. In the remainder of this section we will describe the general procedure employed (details will be omitted from the main text, but can be found in Appendix \ref{section:details_quant} ) and compare our predictions to results from an ensemble of COLA simulations.

We can readily write a formal expression for the variance in these spectra,
\begin{equation}
    \sigma^2_{\mu\nu}(k_i) = \left\langle \widehat{P}^2_{\mu\nu}(k_i) \right\rangle - \left\langle\widehat{P}_{\mu\nu}(k_i)\right\rangle^2,
    \label{eq:variance}
\end{equation}
in which 
\begin{equation}
    \widehat{P}_{\mu\nu}(k_i) = \frac{1}{V_s(k_i)}\int_{k_i} \frac{d^3\bm{k}}{(2\pi)^3} \widehat{P}_{\mu\nu}(\bm{k}) \equiv \left[ \widehat{P}_{\mu\nu}(\bm{k}) \right]_{\Theta(k_i)}, 
\end{equation}
and $V_s(k_i) = 4\pi k_i^3 \Delta(\ln k)$ is the volume of the spherical shell of radius $k_i$ and width $\Delta(\ln k)$, while $\int_{k_i}$ indicates integration over this shell in Fourier space. In order to simplify equation (\ref{eq:variance}) so that we can evaluate it analytically, we will make the approximation that
\begin{equation}
    \left[ \widehat{P}^2_{\mu\nu}(\bm{k}) \right]_{\Theta(k_i)} = \left[\widehat{P}_{\mu\nu}(\bm{k})\right]^2_{\Theta(k_i)},
\end{equation}
and that we are working in the linear regime so that power spectra at different wave vectors will be uncorrelated, that is, $\langle \widehat{P}_{\mu\nu}(\bm{k}) \widehat{P}_{\mu\nu}(\bm{k}') \rangle = 0$ if $\bm{k}\neq \bm{k}'$. Using these approximations, we can write 
\begin{equation}
    \sigma_{\mu\nu}^2(k_i) = \frac{1}{N_{k_i}} \sigma_{\mu\nu}^2(\bm{k}),
\end{equation}
in which $N_{k_i} = \frac{V_s(k_i)}{2V_f}$ is the number of independent Fourier modes falling inside the spherical shell of radius $k_i$, and $V_f=(2\pi)^3/V$ is the volume of one Fourier cell. This expression now makes it simple to compute $\sigma^2_{\mu\nu}(k_i)$ by computing $\sigma^2_{\mu\nu}(\bm{k})$ from the expressions for the fields in appendix \ref{section:advected_fields}, and then dividing them by $N_{k_i}$.

Once established the approximations that we will be using, we now turn to describing how the calculation of $\sigma^2_{\mu\nu}$ takes place. For brevity we will not show the calculations for all the spectra, but merely indicate how one can perform the calculation of $\sigma_{11}(\bm{k})$. Consider equation (\ref{eq:p_11}) for the power spectrum of a fixed simulation, and let us recall that the first and second terms in this equation will not contribute to the variance in the F\&P case. Let us call the third term in this equation $T_3(\bm{k})$,
\begin{equation}
    \begin{split}
        T_3(\bm{k}) = & \frac{V}{4}\int_{\bm{q}_1}\int_{\bm{q}_2} \sqrt{P^L_{\bm{q}_1}P^L_{\bm{k}-\bm{q}_1}P^L_{\bm{q}_2}P^L_{\bm{q}_2-\bm{k}}} \cos\left[ \theta_{\bm{q}_1} + \theta_{\bm{k}-\bm{q}_1} - \theta_{\bm{q}_2} - \theta_{\bm{k}-\bm{q}_2} \right] \\
        & \times F_{ZA}(\bm{q}_1,\bm{k}-\bm{q}_1, \bm{k})F_{ZA}(\bm{q}_2,\bm{k}-\bm{q}_2, \bm{k}),
    \end{split}
\end{equation}
and we can write the variance of $P_{11}$ simply as
\begin{equation}
    \label{eq:var_simple}
    \sigma^2_{11}(\bm{k}) = \langle T_3^2(\bm{k}) \rangle - \langle T_3(\bm{k})\rangle^2.
\end{equation}
To simplify the notation, we will write $T_3(\bm{k})$ as 
\begin{equation}
    T_3(\bm{k}) = \int_{\bm{q}_1} \int_{\bm{q}_2} A(\bm{q}_1, \bm{q}_2, \bm{k}) \cos\left(\Theta\left(\bm{q}_1,\bm{q}_2\right) \right),
\end{equation}
in which we have used
\begin{equation}
    \begin{split}
        A(\bm{q}_1, \bm{q}_2, \bm{k}) = & \frac{V}{4}\sqrt{P_{\bm{q}_1}^L P_{\bm{k}-\bm{q}_1}^LP_{\bm{q}_2}^LP_{\bm{q}_2-\bm{k}}^L}F_{ZA}(\bm{q}_1, \bm{k}-\bm{q}_1)F_{ZA}(\bm{q}_2, \bm{k}-\bm{q}_2)\\
        \Theta(\bm{q}_1, \bm{q}_2, \bm{k}) = & \theta_{\bm{q}_1} + \theta_{\bm{k}-\bm{q}_1} - \theta_{\bm{q}_2} - \theta_{\bm{k}-\bm{q}_2}.
    \end{split}
\end{equation}
Let us begin by computing the first term on the right-hand side of equation (\ref{eq:var_simple}). Using the notation we just defined, we can write it as
\begin{equation}
    \int_{\bm{q}_1, \bm{q}_2, \bm{q}_3, \bm{q}_4}A(\bm{q}_1, \bm{q}_2, \bm{k}) A(\bm{q}_3, \bm{q}_4, \bm{k}) \left\langle \cos\left( \Theta(\bm{q}_1,\bm{q}_2,\bm{k}) \right)\cos\left( \Theta(\bm{q}_3,\bm{q}_4, \bm{k}) \right) \right\rangle.
\end{equation}
Notice that $\Theta$ is a sum of random variables, and is therefore also a random variable uniformly distributed between $0$ and $2\pi$. This implies that if $\Theta(\bm{q}_1, \bm{q}_2, \bm{k})$ is independent of $\Theta(\bm{q}_3, \bm{q}_4, \bm{k})$, we can split the ensemble average of the product into the product of the ensemble averages, which will both be null. Therefore, the only cases for which the above integral will not be zero are when
\begin{equation}
    \cos\left(\Theta(\bm{q}_1,\bm{q}_2,\bm{k})\right) = \cos\left(\Theta(\bm{q}_3,\bm{q}_4,\bm{k})\right),
\end{equation}
as in this case the expression becomes the expected value of the square of the cosine of a random variable, which equals $\frac{1}{2}$, in a well known result. This will be satisfied when one of the following conditions is fulfilled,

\begin{alignat*}{2}
    & \begin{aligned} & \begin{cases}
    \bm{q}_1=\bm{q}_2 & \text{ and } \bm{q}_3=\bm{q}_4, \text{ or } \\
    \bm{q}_1=\bm{k}-\bm{q}_2 & \text{ and } \bm{q}_3=\bm{q}_4, \text{ or } \\
    \bm{q}_1=\bm{q}_2 & \text{ and } \bm{q}_3=\bm{k}-\bm{q}_4, \text{ or } \\
    \bm{q}_1=\bm{k}-\bm{q}_2 & \text{ and } \bm{q}_3=\bm{k}-\bm{q}_4, \text{ or }
  \end{cases}\\
  \MoveEqLeft[-1]
  \end{aligned}
    & \hskip 6em &
  \begin{aligned}
  & \begin{cases}
        \bm{q}_1 = \bm{q}_3 & \text{ and } \bm{q}_2 = \bm{q}_4, \text{ or } \\
        \bm{q}_1 = \bm{q}_4 & \text{ and } \bm{q}_2 = \bm{q}_3, \text{ or } \\
        \bm{q}_1 = \bm{k}-\bm{q}_3 & \text{ and } \bm{q}_2 = \bm{k}-\bm{q}_4, \text{ or } \\
        \bm{q}_1 = \bm{k}-\bm{q}_4 & \text{ and } \bm{q}_2 = \bm{k}-\bm{q}_3, \text{ or } \\
        \bm{q}_1 = \bm{q}_3 & \text{ and } \bm{q}_2 = \bm{k}-\bm{q}_4, \text{ or } \\
        \bm{q}_1 = \bm{k}-\bm{q}_3 & \text{ and } \bm{q}_2 = \bm{q}_4, \text{ or }\\
        \bm{q}_1 = \bm{q}_4 & \text{ and } \bm{q}_2 = \bm{k}-\bm{q}_3, \text{ or }\\
        \bm{q}_1 = \bm{k}-\bm{q}_4 & \text{ and } \bm{q}_2 = \bm{q}_3.
  \end{cases} \\[3.3ex]
  \MoveEqLeft[-1]
  \end{aligned}
\end{alignat*}
Each of these cases will give a contribution, e.g.: for the first case of the block on the right we substitute the ensemble average of the cosines by the expression
\begin{equation}
    \underbrace{\frac{1}{2}}_{\langle\cos^2\rangle} \times \underbrace{\frac{1}{L^3}\delta^D(\bm{q}_1-\bm{q}_3)}_{\bm{q}_1=\bm{q}_3}\times\underbrace{\frac{1}{L^3}\delta^D(\bm{q}_2-\bm{q}_4)}_{\bm{q}_2=\bm{q}_4}.
\end{equation}
We have divided these cases into two blocks because the block on the left will give contributions which will precisely cancel the second term in the RHS of equation (\ref{eq:var_simple}), and contributions to the variance will only come from the block to the right. Fortunately, for the case of $\sigma_{11}^2$ all of the cases will contribute the same value, and thus we can write the expression for the variance, having already integrated over the Dirac deltas,
\begin{equation}
    \sigma^2_{11}(\bm{k}) \approx \int_{\bm{q}} P(q)P(|\bm{k}-\bm{q}|) F^2_{ZA}(\bm{q},\bm{k}-\bm{q}),
\end{equation}
a remarkable simplification compared to the expressions from which we began.

The results obtained from these calculations are shown in figure (\ref{fig:cola_var}), along with the variances computed from the ensemble of COLA simulations. We can see that our calculations are able to reproduce well most of the variances up to scales of $k\sim 0.2[\text{Mpc}/h]$; those which are not well reproduced such as $\sigma_{11}, \sigma_{1\delta}$ and $\sigma_{\delta\delta}$ are the ones which have their leading-order and next to leading order contributions cancelled, and therefore, the calculation of their variances becomes quite complex, requiring us to go to higher orders in perturbation theory to have very precise predictions -- a task already beyond the scope of the current work. 

\section{Optimal Phase-Fields}
\label{section:ph_field}

The previous section has demonstrated the power of F\&P in reducing statistical noise in simulations, and thus its great potential for building simulation-based theoretical models for galaxy clustering. We have seen, however, that these techniques do not reduce the variance in all basis spectra -- as a direct consequence, for certain combinations of values of the bias parameters, $\sigma_{gg}$ will be dominated by contributions coming from $\sigma_{\delta^2\delta^2}, \sigma_{\delta^2s^2}$ and $\sigma_{s^2s^2}$, precisely the terms which get no variance reduction. The question that poses itself naturally is whether there is a way to reduce these terms as well.

Without modifying the underlying physics, the freedom we have in an N-Body simulation is essentially that of changing the initial conditions -- amplitudes and phases of the Fourier modes $\delta^L(\bm{k})$. Using F\&P the amplitudes are no longer free after we fix them to be exactly the square root of the power spectrum, but the phases remain unconstrained. One could think of these phases as parameters to be adjusted, and try to fit them so that the spectra measured from the simulation match their ensemble averages; this would of course have a prohibitive cost if for every point to be tested in this extremely high-dimensional space we should have to run a complete simulation and then compute its power spectra. This prompts us to look for an approximation to the power spectra of the fully evolved simulations such that it can be quickly calculated, and captures the statistical properties of these power spectra, namely their deviations from the ensemble average.

At linear scales it is known that the growth of structures proceeds independently between different scales $k\neq k'$. This means that there is little mode mixing at these scales, so that the statistical properties of these modes in the initial conditions will be essentially preserved in the final output of a simulation. Therefore, the level of noise of e.g.: $P_{\delta^2\delta^2}$ computed from the initial conditions of a simulation $\delta^{IC}(\bm{\theta})$ will be very similar to the level of noise in $P_{\delta^2\delta^2}$ computed from the fully evolved field, provided we compare them at sufficiently large scales. This gives us an approximation with the characteristics we were looking for: computing the power spectra of the initial conditions is very fast, and approximately captures the statistical properties of the power spectra of the fully evolved field.

\begin{figure}[t!]
    \centering
    \includegraphics[width=\textwidth]{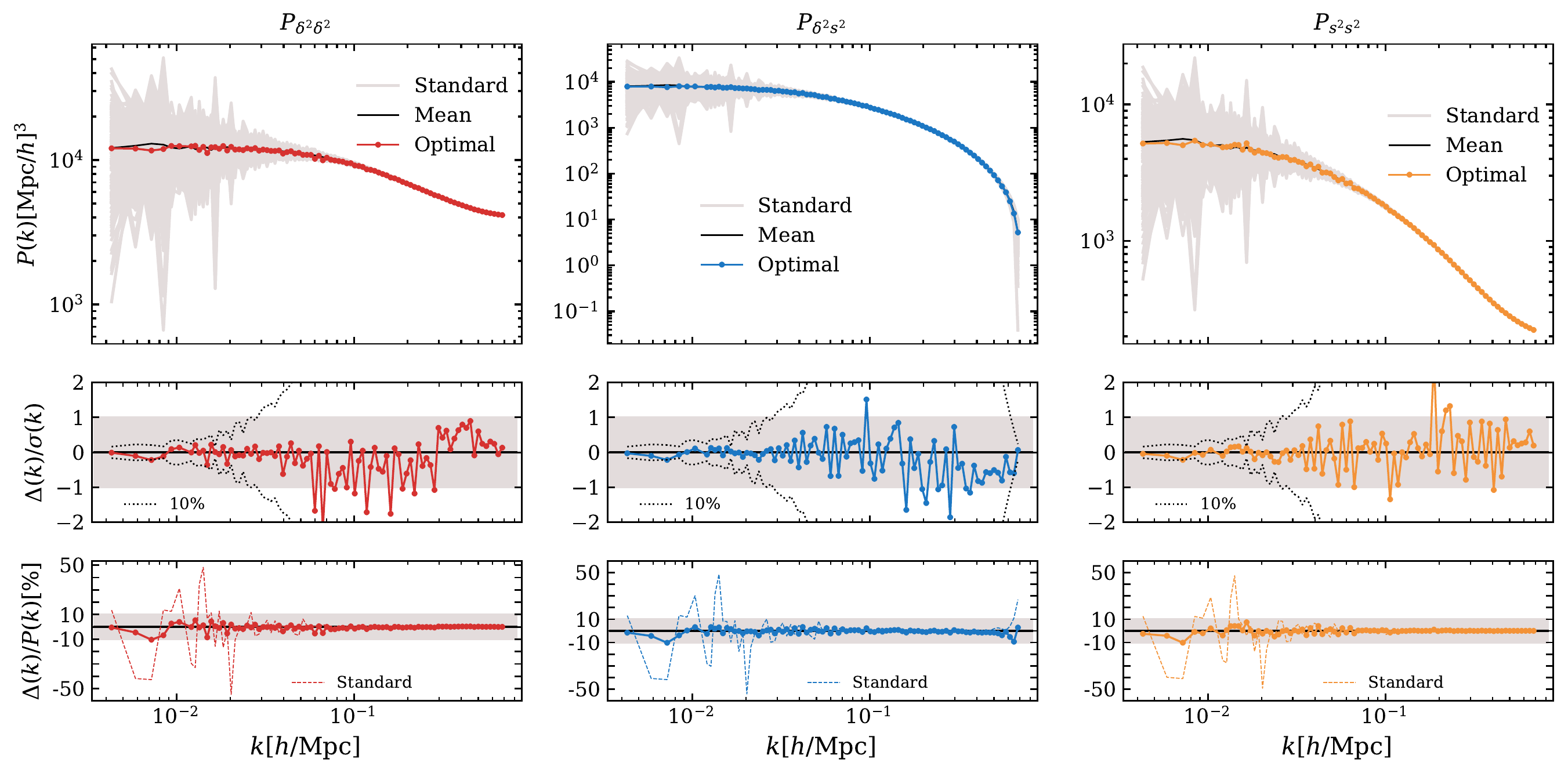}
    \caption{In this figure we compare the deviations of the optimal spectrum from the ensemble average to those from a random realization. In the first row the solid line with circular markers shows the optimal spectrum, the gray lines show 200 spectra from random realizations, and the black line shows their mean. In the second row the solid line with circular markers shows the deviation of the optimal spectrum from the ensemble average, divided by the standard-deviation in the set of 200 random spectra; the black dotted line shows bands corresponding to a $10\%$ deviation from the mean. In the final row the solid line with circular markers represents the fractional difference of the optimal spectrum to the ensemble mean, and the dashed line shows the same quantity for a random realization.}
    \label{fig:optimal_spec}
\end{figure}

Having found a sufficiently fast approximation to the power spectra of a fully evolved simulation, we can now use it to find the optimal phase-field that reduces the distance of the relevant power spectra to their ensemble averages. Notice that the ensemble average can be easily computed up to the desired precision, simply by generating as many realizations of the initial conditions as needed, computing their power spectra and averaging over them. Further explaining and systematizing the optimization procedure, we can divide it into three steps:

\begin{enumerate}
    \item \textbf{Generating or Receiving ICs.} First the amplitudes of the Fourier modes are generated $|\delta_L(\bm{k})|=A(k) = \sqrt{L^3P^L(k)}$, and the phases are either generated randomly or received from an iteration of this whole procedure, allowing us to form the initial density field $\delta^L(\bm{k}) = A(k)e^{i\theta_i(\bm{k})}$;
    
    \item \textbf{Computing the Loss-Function.} We compute $P_{\delta^2\delta^2}(\bm{\theta}_i)$, $P_{\delta^2 s^2}(\bm{\theta}_i)$ and $P_{s^2s^2}(\bm{\theta}_i)$ taking $\delta^L(\bm{k})$ as input, and use this to calculate a loss-function that measures the distances of these spectra from their ensemble averages. This is done by taking advantage of the Python library JAX \cite{jax2018github}; functions implemented using this library can be automatically differentiated, allowing one to evaluate their gradients without having to write any additional code; this feature will be very important for the next step, at which the gradient of this function will be used inside a minimization algorithm.
    
    \item \textbf{Stepping Minimization Algorithm.} The value of the loss-function and its derivative with respect to the phases is fed to the SciPy implementation of the L-BFGS-B algorithm \cite{lbfgs, scipy_2020}, that produces an updated phase field $\bm{\theta}_{i+1}$ with a smaller value of the loss-function. Steps 1-3 are repeated until the loss-function is reduced by a specified factor, at which point one says the algorithm has converged, and outputs an optimal phase-field $\bm{\theta}_{opt}$.
    
\end{enumerate}

Having carried out this process, one can run a full simulation using the optimal phase-field as input and compare the power-spectra measured from it to those from a standard simulation. Figure (\ref{fig:optimal_spec}) shows precisely this comparison for the three spectra of interest; one can see how the deviations from the ensemble average are significantly reduced at large scales, corresponding to a small fraction of the standard deviation in simulations with random phase-fields. Furthermore, these deviations are restricted to less than $10\%$ at all scales -- this is particularly impressive for large scales, at which deviations in standard simulations can easily reach $40\%$. 

This optimization algorithm provides very distinct results depending on the given initial phase-field; taking advantage of that, we ran it using different initial conditions to find five independent optimal phase-fields. From the simulations run with these initial conditions one can then estimate the power spectra and compute their variance and compare it to the result for usual F\&P. This comparison can be seen in figure (\ref{fig:var_optimal_comp_fits}) of appendix \ref{section:optimization_details} and shows a reduction by a roughly a factor of $10$ in the variances of $P_{\delta^2\delta^2}$, $P_{\delta^2s^2}$ and $P_{s^2s^2}$.

One would hope that this optimization procedure could reveal general properties of the optimal phase-fields responsible for reducing the noise. However, we have investigated the probability density function of the phases and its power spectrum, but did not find significant differences to the same properties computed for random realizations. This can be most likely attributed to the fact that this optimization algorithm has the freedom to fit a very large number of free parameters -- the values of the initial phases -- and therefore the solution that we find is a very specific one. Indeed, if one tries to use this same phase-field to initialize a simulation at a different cosmology or with a different box size, the optimal properties are quickly lost. 

\section{EUCLID Forecast}
\label{section:euclid_forecast}

\begin{figure}[tbp]
\centering 
\includegraphics[width=0.95\textwidth]{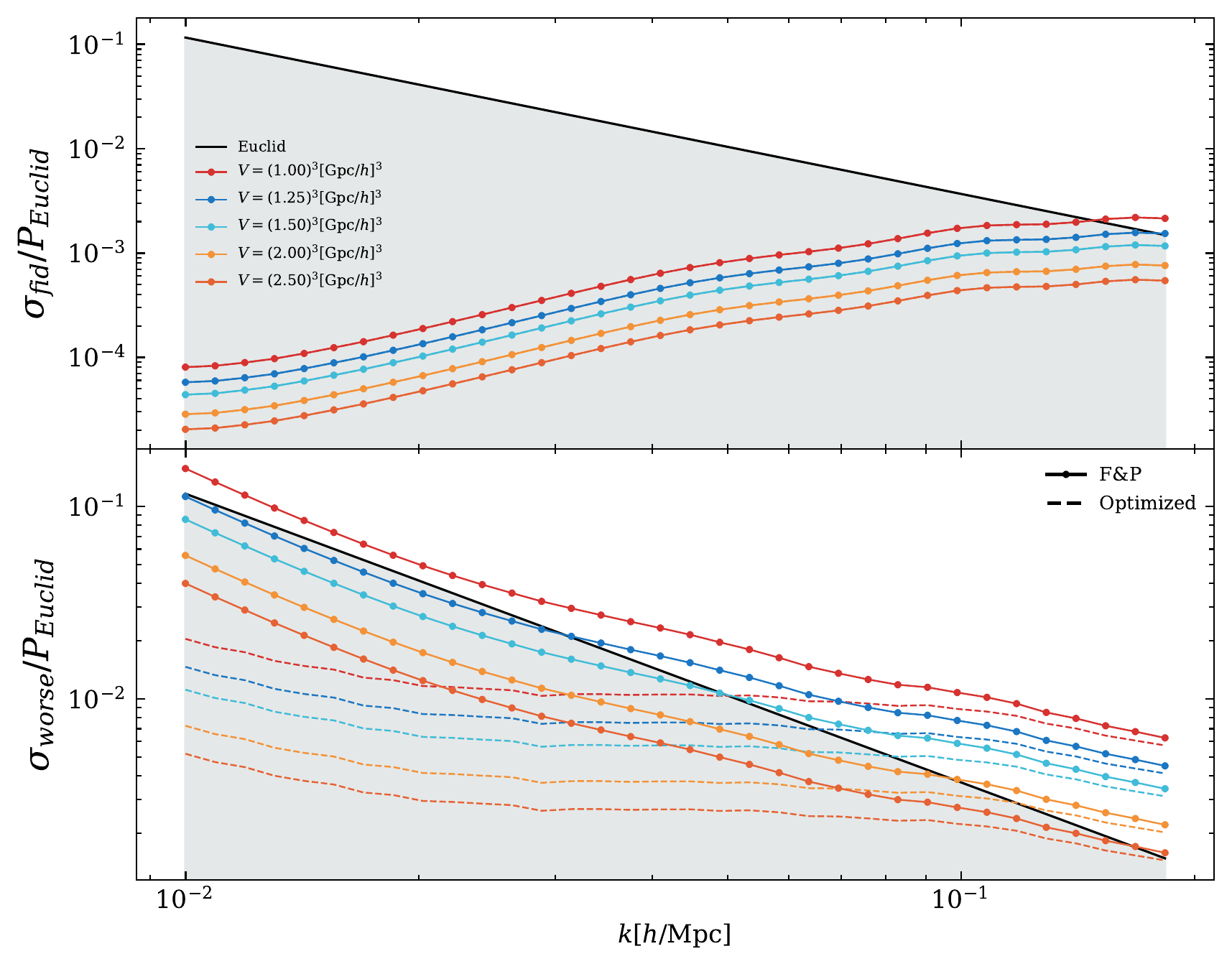}
\caption{This figure shows the comparison of the model error for several simulation sizes to the expected error made by Euclid in measuring power spectra. In the top panel we show these quantities computed assuming $b_1^L=0.38$ and using the coevolution relations in \cite{zennaro2021priors} to obtain values for the higher-order bias parameters. In the bottom panel we show these quantities computed assuming the same linear bias, but assigning to $b_{\delta^2}$ and $b_{s^2}$ the largest values allowed for these bias parameters according to \cite{zennaro2021priors}. }
\label{fig:euclid-error}
\end{figure}

The error in hybrid Lagrangian bias expansion models of large-scale structure will have multiple sources, depending on each particular implementation. One main source will be related to the interpolation scheme used to allow access to the model in any required cosmology; another source is associated to the algorithms employed in carrying out simulations; finally, there is the error coming from the statistical noise of the simulation itself. The latter arises because of the finite volume of the simulations, and is precisely the one which can be reduced through F\&P techniques, or through the optimization procedure described above.

This statistical noise in the dark-matter only simulations will translate itself to a source of noise in the galaxy power spectrum; we can write this relationship using once again the bias expansion
\begin{equation}
\begin{split}
    \sigma^2_{gg} = \sum_{\alpha,\beta,\mu,\nu \in [1, \delta, \delta^2, s^2, \nabla^2\delta]} b_\alpha b_\beta b_\mu b_\nu \sigma^2_{\alpha\beta,\mu\nu},
    \label{eq:bias_exp_var}
\end{split}
\end{equation}
and in this expression we have slightly changed our notation for the variances. This was necessary because our previous notation did not allow us to represent cross-variances between different spectra, which we now write as
\begin{equation}
    \sigma_{\alpha\beta,\mu\nu}^2 = \langle (P_{\alpha\beta}-\bar{P}_{\alpha\beta})(P_{\mu\nu}-\bar{P}_{\mu\nu}) \rangle. 
\end{equation}
These cross-variances can give quite large contributions to the galaxy spectrum variance, which represents an issue since our calculations are not capable of giving predictions for them; however, as we have seen, one can divide these spectra into three groups, namely $\left\{P_{11}, P_{1\delta}, P_{\delta\delta}\right\}$, $\left\{P_{1\delta^2}, P_{1s^2}, P_{\delta\delta^2}, P_{\delta s^2}\right\}$ and $\left\{P_{\delta^2\delta^2}, P_{\delta^2s^2}, P_{s^2s^2}\right\}$ such that the spectra belonging to a same group have very similar noise properties. Therefore, we will assume that if the spectra $P_{\alpha\beta}$ and $P_{\mu\nu}$ belong to the same group, their cross-variance will be given by
\begin{equation}
    \sigma_{\alpha\beta,\mu\nu} = \sqrt{\sigma_{\alpha\beta} \sigma_{\mu\nu}},    
\end{equation}
and if they belong to different groups, 
\begin{equation}
    \sigma_{\alpha\beta,\mu\nu} = 0.
\end{equation}
Using these approximations and our calculations of the variances $\sigma_{\mu\nu}$, $\mu,\nu\in[1, \delta, \delta^2, s^2, \nabla^2\delta]$ for F\&P simulations we can thus compute the contribution of the statistical noise in simulations to the final error in the galaxy power-spectrum, and even forecast the size of the simulations one should run to have this term below the expected measurement error in EUCLID galaxy power spectra.

\begin{figure}[t!]
    \centering
    \includegraphics[width=\textwidth]{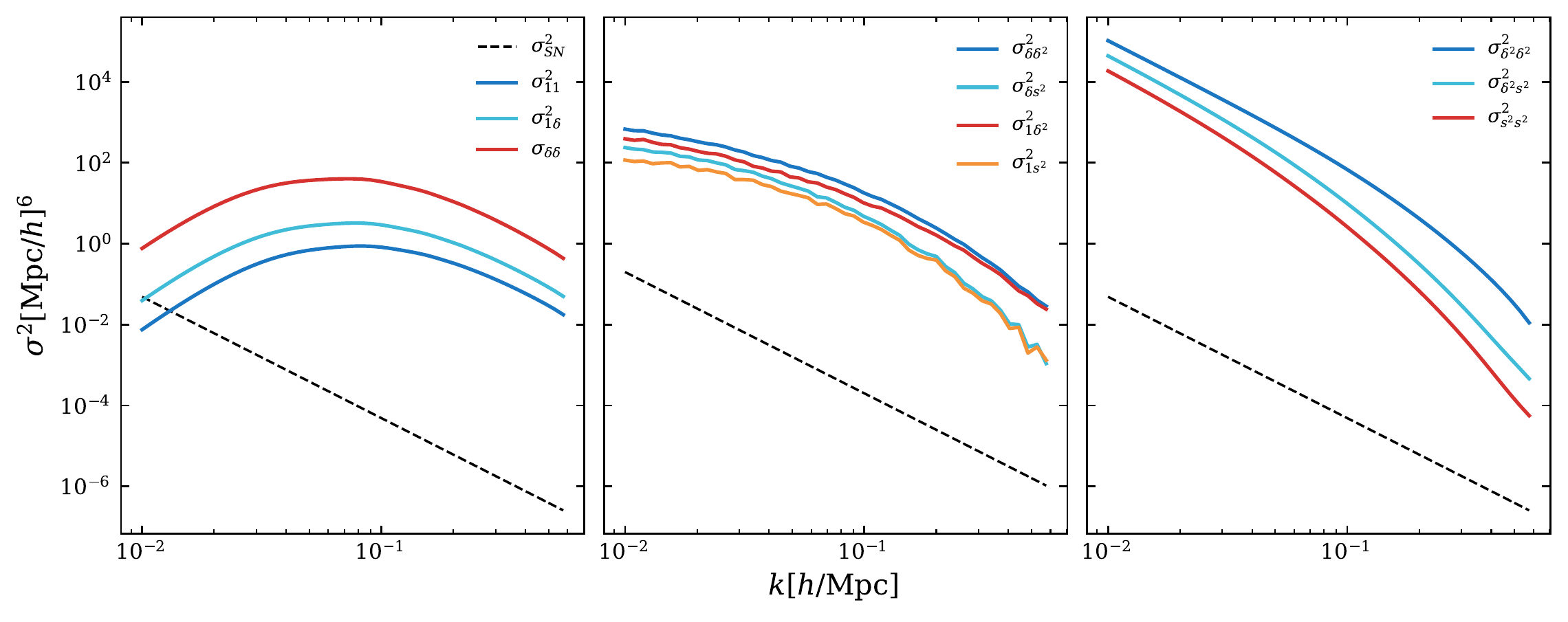}
    \caption{Figure showing the amplitudes of the variances for the auto and cross-spectra entering the bias expansion for the galaxy spectrum. Notice that these terms will contribute to equation (\ref{eq:bias_exp_var}) modulated by their respective bias parameters, which can greatly suppress or amplify their importance in the final error. The black dashed line shows the contribution to the variance coming from shot-noise in our dark-matter only simulation.}
    \label{fig:euclid_break}
\end{figure}

The calculation of the measurement error for galaxy power-spectra in an EUCLID-like survey was done using the parameters shown in table (\ref{Tab:pars_euclid}), and the expression for the variance in a random Gaussian field
\begin{equation}
    \sigma^2_{\text{E}} = \frac{2}{N_k}P_{gg}^2,
\end{equation}
in which $N_k = 4\pi k^3 \Delta(\ln k) V_{\text{E}} / (2\pi)^3$ is the number of Fourier modes falling the spherical shell of radius $k$, and the power spectrum is given by 
\begin{equation}
    P_{gg} = \sum_{i,j \in [1, \delta, \delta^2, s^2, \nabla^2\delta]} b_ib_j P_{ij} + \frac{1}{\bar{n}}.
\end{equation}
We have considered two separate cases for the values of the bias parameters. In the first case we do a fiducial analysis, considering the bias parameters which seem most likely to be found by the EUCLID survey; for that, we fixed the linear Lagrangian bias to its expected value from forecasts, $b_1^L = 0.38$ \cite{euclid_2018_amendola}, and the values of the other bias parameters were obtained from the co-evolution relations for galaxies found in \cite{zennaro2021priors}. The second scenario we have considered is one in which $b_{\delta^2}$ and $b_{s^2}$ take the maximum values of the intervals over which \cite{zennaro2021priors} observed them to vary; the motivation for this is that in the fiducial case the contributions coming from $\sigma_{\delta^2\delta^2}$, $\sigma_{\delta^2s^2}$ and $\sigma_{s^2s^2}$ are highly suppressed due to the very small values taken by their accompanying bias parameters, despite their sheer values being much larger than for the other spectra, as can be seen in figure (\ref{fig:euclid_break}). Therefore, relatively small variations in these bias parameters could lead to very different scenarios in which the model error is actually dominated by the noise in these spectra; the second scenario thus serves to probe such regimes.

From figure (\ref{fig:euclid-error}) one can see that a F\&P simulation with a volume $V \sim 2 [$Gpc$/h]^3$ would be sufficient to have the statistical errors smaller than what is expected for EUCLID, in the case of the fiducial bias parameters. This shows the great power of the F\&P technique, as this volume corresponds to roughly $6\%$ of the assumed volume for EUCLID; using a simulation with usual initial conditions would require a volume at least as large as that of the survey. For the second set of biases the error becomes dominated by $\sigma_{\delta^2\delta^2}$, $\sigma_{\delta^2s^2}$ and $\sigma_{s^2s^2}$, and the necessary volume for the statistical error to be below EUCLID errors is around $V\sim 16 [$Gpc$/h]^3$; the dashed lines in this figure show how in this case our optimization procedure helps reduce by a factor of almost 10 the error in the final galaxy spectrum at the largest scales. In summary, these results show that controlling this source of error is within reach, and that obtaining a model with the precision required for EUCLID analysis will depend mostly on reducing emulation and rescaling errors, as these are in the range $1\% \sim 3\%$, and thus we can see from figure (\ref{fig:euclid-error}) that they would dominate over the statistical noise.

\section{Conclusion}
\label{section:conclusion}

With the analyses presented in this we have demonstrated the power of F\&P simulations for building hybrid Lagrangian bias expansion models of LSS. We have shown that the method does not introduce any biases to the relevant spectra, while significantly reducing the variance in the final galaxy spectrum, and thus easing the computational tasks of running simulations large enough to have the required precision to analyse upcoming survey data. This establishes this technique firmly as a foremost method to accessing accurate and precise predictions of LSS summary statistics.

We have also explored the method’s deficiencies, namely the fact that it does not reduce the variance for all of the basis spectra. We provided a clear analytical explanation for these observations based on LPT; the same formalism also allowed us to obtain predictions for the variances of these basis spectra in F\&P simulations. These results should be fundamental for future analyses that use hybrid Lagrangian bias expansion models, as they are a direct and extremely flexible way to obtain the error being made in the model, and this is expected to have a sizeable importance with the continuous reduction of experimental errors.

Using the analytical predictions for the variances in the basis spectra we have also been able to develop an understanding of the numerical challenges that have to be overcome to make hybrid Lagrangian bias expansion models precise enough for the analysis of EUCLID data. Our results clearly indicate that, assuming the real biases to be between the fiducial and worst-case scenarios, simulations with $V \sim 8 [$Gpc$/h]^3$ should be sufficient to make the statistical noise in the model smaller than EUCLID errors; this volume is well within reach of current supercomputing facilities. Therefore, other sources of error such as those from interpolation schemes or from techniques to scale existing simulations to new cosmologies, are the ones that will need to be mitigated.

Finally, it was also possible to go beyond the limitations of the method, reducing the variance in terms that remain otherwise unaltered, namely $\sigma_{\delta^2\delta^2}$, $\sigma_{\delta^2 s^2}$ and $\sigma_{s^2s^2}$. For the simulation size and resolution used, the method has proved to be unbiased, and the reduction in the variances was substantial, reason enough to consider it as a good option to reduce statistical noise rather than simply running larger simulations. It also highlights the influence of the initial phase-field in the derived $n$-point functions, showing that there exist a large number of initial configurations that lead to results arbitrarily close to the ensemble average.

\appendix
\section{Advected Lagrangian Fields}
\label{section:advected_fields}

In this appendix we show the results of applying equation (\ref{eq:advect}) to each of the $F(\bm{q})\in\left[ 1, \delta, \delta^2, s^2, \nabla^2\delta \right]$. It is convenient to define a kernel which will appear many times in the following calculations,
\begin{equation}
    F_{ZA}(\bm{q}_1, \bm{q}_2) = 1 + \frac{\bm{q}_1\cdot\bm{q_2}}{q_1q_2}\left( \frac{q_1}{q_2} + \frac{q_2}{q_1} \right) + \frac{(\bm{q}_1\cdot\bm{q}_2)^2}{q_1^2q_2^2}.
\end{equation}

\subsection{$\delta_1$}

\begin{equation}
    \delta_1(\bm{k}) = \delta_{\bm{k}}^L+\frac{1}{2}\int_{\bm{q}_1} F_{ZA}(\bm{q}_1,\bm{k}-\bm{q}_1)\delta^L_{\bm{q}_1}\delta^L_{\bm{k}-\bm{q}_1} + \mathcal{O}(\delta^3)
    \label{eq:delta_1}
\end{equation}

\subsection{$\delta_{\delta}$}

\begin{equation}
    \delta_{\delta}(\bm{k}) = \delta_{\bm{k}}^L + \int_{\bm{q}_1}\delta^L_{\bm{q}_1}\delta^L_{\bm{k}-\bm{q}_1}\frac{\bm{k}\cdot(\bm{k}-\bm{q}_1)}{|\bm{k}-\bm{q}_1|^2} + \frac{1}{2}\int_{\bm{q}_1,\bm{q}_2}\delta^L_{\bm{q}_1} \delta^L_{\bm{q}_2} \delta^L_{\bm{k}-\bm{q}_{12}} F_{ZA}(\bm{q}_2,\bm{k}-\bm{q}_{12}) + \mathcal{O}(\delta^4)
    \label{eq:delta_delta}
\end{equation}

\subsection{$\delta_{\delta^2}$}

\begin{equation}
    \delta_{\delta^2}(\bm{k}) = \int_{\bm{q}_1} \delta^L_{\bm{q}_1}\delta^L_{\bm{k}-\bm{q}_1} + \int_{\bm{q}_1,\bm{q}_2}\delta^L_{\bm{q}_1}\delta^L_{\bm{q}_2}\delta^L_{\bm{k}-\bm{q}_{12}}\frac{\bm{k}\cdot(\bm{k}-\bm{q}_{12})}{|\bm{k}-\bm{q}_{12}|^2}+\mathcal{O}(\delta^4) 
    \label{eq:delta_delta_2}
\end{equation}

\subsection{$\delta_{s^2}$}

\begin{equation}
    \delta_{s^2}(\bm{k}) = \int_{\bm{q}_1}\delta^L_{\bm{q}_1}\delta^L_{\bm{k}-\bm{q}_1} S_2(\bm{q}_1,\bm{k}-\bm{q}_1) + \int_{\bm{q}_1,\bm{q}_2}\frac{\bm{k}\cdot(\bm{k}-\bm{q}_{12})}{|\bm{k}-\bm{q}_{12}|^2}S_2(\bm{q}_1,\bm{q}_2)\delta^L_{\bm{q}_1}\delta^L_{\bm{q}_2}\delta^L_{\bm{k}-\bm{q}_{12}} + \mathcal{O}(\delta^4)
    \label{eq:delta_s_2}
\end{equation}

\subsection{$\delta_{\nabla^2\delta}$}

\begin{equation}
    \delta_{\nabla^2\delta}(\bm{k}) = k^2\delta^L_{\bm{k}} + \int_{\bm{q}_1}q_1^2\delta^L_{\bm{q}_1}\delta^L_{\bm{k}-\bm{q}_1}\frac{\bm{k}\cdot(\bm{k}-\bm{q}_1)}{|\bm{k}-\bm{q}_1|^2} + \frac{1}{2}\int_{\bm{q}_1,\bm{q}_2}q_1^2\delta^L_{\bm{q}_1} \delta^L_{\bm{q}_2} \delta^L_{\bm{k}-\bm{q}_{12}} F_{ZA}(\bm{q}_2,\bm{k}-\bm{q}_{12}) + \mathcal{O}(\delta^4)
    \label{eq:delta_nabla_2_delta}
\end{equation}

\section{Further Details on Quantitative Predictions}
\label{section:details_quant}

In this appendix we will provide further details on the calculations sketched in section 5.5 . Because we have already described the calculation of $\sigma_{11}(k)$ in the main text, we will detail here the calculation of $\sigma_{1\delta}(k)$. We omit the calculations for the other spectra, since they are completely analogous. 

\subsection{$\sigma_{1\delta}$}

Using the expressions for the advected fields, obtained in appendix \ref{section:advected_fields}, we can readily write an expression for the $P_{1\delta}$ spectrum
\begin{equation}
    \begin{split}
        \frac{1}{L^3} \delta_{\delta}(\bm{k})\delta_1(-\bm{k}) = & P^L(k) + \mathcal{O}(3) \\
        & + \frac{1}{2} \int_{\bm{q}_1,\bm{q}_2} \sqrt{P^L_{\bm{q}_1}P^L_{\bm{k}-\bm{q}_1}P^L_{\bm{q}_2}P^L_{\bm{k}-\bm{q}_2}} \frac{\bm{k}\cdot(\bm{k}-\bm{q}_1)}{|\bm{k}-\bm{q}_1|^2} F_{ZA}(\bm{q}_2,\bm{k}-\bm{q}_{2}) \\
        & \times \cos(\theta_{q_1} + \theta_{\bm{k}-\bm{q}_1} -\theta_{q_2} - \theta_{\bm{k}-\bm{q}_2})
    \end{split}
\end{equation}
in which we have not explicitly written the third-order terms because we already know they will be cancelled by pairing, and are thus irrelevant to the current calculation. The only term contributing to the variance will be the fourth-order one,
\begin{equation}
    \begin{split}
        T_1(\bm{k}) = & \frac{1}{2} \int_{\bm{q}_1,\bm{q}_2} \sqrt{P^L_{\bm{q}_1}P^L_{\bm{k}-\bm{q}_1}P^L_{\bm{q}_2}P^L_{\bm{k}-\bm{q}_2}} \frac{\bm{k}\cdot(\bm{k}-\bm{q}_1)}{|\bm{k}-\bm{q}_1|^2} F_{ZA}(\bm{q}_2,\bm{k}-\bm{q}_{2}) \cos(\theta_{q_1} + \theta_{\bm{k}-\bm{q}_1} -\theta_{q_2} - \theta_{\bm{k}-\bm{q}_2})\\
         = & \int_{\bm{q}_1, \bm{q}_2} A(\bm{q}_1, \bm{q}_2, \bm{k}) \cos\left(\Theta(\bm{q}_1, \bm{q}_2, \bm{k})\right),
    \end{split}
\end{equation}
in which 
\begin{equation}
    \begin{split}
        A(\bm{q}_1, \bm{q}_2, \bm{k}) & = \sqrt{P^L_{\bm{q}_1}P^L_{\bm{k}-\bm{q}_1}P^L_{\bm{q}_2}P^L_{\bm{k}-\bm{q}_2}} \frac{\bm{k}\cdot(\bm{k}-\bm{q}_1)}{|\bm{k}-\bm{q}_1|^2} F_{ZA}(\bm{q}_2, \bm{k}-\bm{q}_2) \\
        \Theta(\bm{q}_1, \bm{q}_2, \bm{k}) & = \theta_{\bm{q}_1} + \theta_{\bm{k}-\bm{q}_1} - \theta_{\bm{q}_2} - \theta_{\bm{k}-\bm{q}_2}
    \end{split}
\end{equation}
We must compute the expression
\begin{equation}
    \langle T_1^2(\bm{k}) \rangle - \langle T_1(\bm{k})\rangle^2,
    \label{eq:var_new}
\end{equation}
and we begin by evaluating the first term in this expression, which we rewrite as 
\begin{equation}
    \begin{split}
        \langle T_1^2(\bm{k}) \rangle & = \int_{\bm{q}_1, \bm{q}_2, \bm{q}_3, \bm{q}_4}A(\bm{q}_1, \bm{q}_2, \bm{k}) A(\bm{q}_3, \bm{q}_4, \bm{k}) \left\langle \cos\left( \Theta(\bm{q}_1,\bm{q}_2,\bm{k}) \right)\cos\left( \Theta(\bm{q}_3,\bm{q}_4, \bm{k}) \right) \right\rangle.
    \end{split}
\end{equation}
Once again, we see that the ensemble average in the expression above will give a non-zero contribution only when we have
\begin{equation}
    \cos\left( \Theta(\bm{q}_1,\bm{q}_2,\bm{k}) \right) = \cos\left( \Theta(\bm{q}_3,\bm{q}_4, \bm{k}) \right),
\end{equation}
which implies 
\begin{equation}
    \theta_{\bm{q}_1} + \theta_{\bm{k}-\bm{q}_1} - \theta_{\bm{q}_2} - \theta_{\bm{k}-\bm{q}_2} = \theta_{\bm{q}_3} + \theta_{\bm{k}-\bm{q}_3} - \theta_{\bm{q}_4} - \theta_{\bm{k}-\bm{q}_4}. 
\end{equation}
This will be satisfied if one of the following conditions is satisfied
\begin{alignat*}{2}
    & \begin{aligned} & \begin{cases}
    \bm{q}_1=\bm{q}_2 & \text{ and } \bm{q}_3=\bm{q}_4, \text{ or } \\
    \bm{q}_1=\bm{k}-\bm{q}_2 & \text{ and } \bm{q}_3=\bm{q}_4, \text{ or } \\
    \bm{q}_1=\bm{q}_2 & \text{ and } \bm{q}_3=\bm{k}-\bm{q}_4, \text{ or } \\
    \bm{q}_1=\bm{k}-\bm{q}_2 & \text{ and } \bm{q}_3=\bm{k}-\bm{q}_4, \text{ or }
  \end{cases}\\
  \MoveEqLeft[-1]
  \end{aligned}
    & \hskip 6em &
  \begin{aligned}
  & \begin{cases}
        \bm{q}_1 = \bm{q}_3 & \text{ and } \bm{q}_2 = \bm{q}_4, \text{ or } \\
        \bm{q}_1 = \bm{q}_4 & \text{ and } \bm{q}_2 = \bm{q}_3, \text{ or } \\
        \bm{q}_1 = \bm{k}-\bm{q}_3 & \text{ and } \bm{q}_2 = \bm{k}-\bm{q}_4, \text{ or } \\
        \bm{q}_1 = \bm{k}-\bm{q}_4 & \text{ and } \bm{q}_2 = \bm{k}-\bm{q}_3, \text{ or } \\
        \bm{q}_1 = \bm{q}_3 & \text{ and } \bm{q}_2 = \bm{k}-\bm{q}_4, \text{ or } \\
        \bm{q}_1 = \bm{k}-\bm{q}_3 & \text{ and } \bm{q}_2 = \bm{q}_4, \text{ or }\\
        \bm{q}_1 = \bm{q}_4 & \text{ and } \bm{q}_2 = \bm{k}-\bm{q}_3, \text{ or }\\
        \bm{q}_1 = \bm{k}-\bm{q}_4 & \text{ and } \bm{q}_2 = \bm{q}_3.
  \end{cases} \\[3.3ex]
  \MoveEqLeft[-1]
  \end{aligned}
\end{alignat*}
Each of these terms will give us a contribution, and once again we will not explicitly write the terms in the left block, because we affirm they will cancel the second term appearing in expression (\ref{eq:var_new}). The contributions coming from the conditions in the right block will be 
\begin{equation}
    \begin{split}
        A_1 = &  \frac{1}{8} \int_{\bm{q}_1} P^L_{q_1}P^L_{|\bm{k}-\bm{q}_1|} \left(\frac{\bm{k}\cdot(\bm{k}-\bm{q}_1)}{|\bm{k}-\bm{q}_1|^2}\right)^2 \int_{\bm{q}_2} P^L_{q_2}P^L_{|\bm{k}-\bm{q}_2|} F_{ZA}^2(\bm{q}_2, \bm{k}-\bm{q}_2) \\
        A_2 = & \frac{1}{8} \left[\int_{\bm{q}} P^L_{q}P^L_{|\bm{k}-\bm{q}|} \frac{\bm{k}\cdot(\bm{k}-\bm{q})}{|\bm{k}-\bm{q}|^2}F_{ZA}(\bm{q},\bm{k}-\bm{q}) \right]^2\\
        A_3 = & \frac{1}{8} \int_{\bm{q}_1} P^L_{q_1}P^L_{|\bm{k}-\bm{q}_1|} \frac{\bm{k}\cdot(\bm{k}-\bm{q}_1)}{|\bm{k}-\bm{q}_1|^2} \frac{\bm{k}\cdot\bm{q}_1}{q_1^2} \int_{\bm{q}_2} P^L_{q_2} P^L_{|\bm{k}-\bm{q}_2|} F_{ZA}^2(\bm{q}_2,\bm{k}-\bm{q}_2)\\
        A_4 = & \frac{1}{8} \left[\int_{\bm{q}} P^L_{q}P^L_{|\bm{k}-\bm{q}|} \frac{\bm{k}\cdot(\bm{k}-\bm{q})}{|\bm{k}-\bm{q}|^2}F_{ZA}(\bm{q},\bm{k}-\bm{q}) \right]^2\\
        A_5 = &  \frac{1}{8} \int_{\bm{q}_1} P^L_{q_1}P^L_{|\bm{k}-\bm{q}_1|} \left(\frac{\bm{k}\cdot(\bm{k}-\bm{q}_1)}{|\bm{k}-\bm{q}_1|^2}\right)^2 \int_{\bm{q}_2} P^L_{q_2}P^L_{|\bm{k}-\bm{q}_2|} F_{ZA}^2(\bm{q}_2, \bm{k}-\bm{q}_2) \\
        A_6 = & \frac{1}{8} \int_{\bm{q}_1} P^L_{q_1}P^L_{|\bm{k}-\bm{q}_1|} \frac{\bm{k}\cdot(\bm{k}-\bm{q}_1)}{|\bm{k}-\bm{q}_1|^2} \frac{\bm{k}\cdot\bm{q}_1}{q_1^2} \int_{\bm{q}_2} P^L_{q_2} P^L_{|\bm{k}-\bm{q}_2|} F_{ZA}^2(\bm{q}_2,\bm{k}-\bm{q}_2)\\
        A_7 = & \frac{1}{8} \left[\int_{\bm{q}} P^L_{q}P^L_{|\bm{k}-\bm{q}|} \frac{\bm{k}\cdot(\bm{k}-\bm{q})}{|\bm{k}-\bm{q}|^2}F_{ZA}(\bm{q},\bm{k}-\bm{q}) \right]^2\\
        A_8 = & \frac{1}{8} \left[\int_{\bm{q}} P^L_{q}P^L_{|\bm{k}-\bm{q}|} \frac{\bm{k}\cdot(\bm{k}-\bm{q})}{|\bm{k}-\bm{q}|^2}F_{ZA}(\bm{q},\bm{k}-\bm{q}) \right]^2.
    \end{split}
\end{equation}
Many of these contributions are equal, and summing them up, we obtain that the variance of $P_{1\delta}$ in a fixed and paired simulation is given approximately by
\begin{equation}
    \begin{split}
            \sigma_{1\delta}^2(\bm{k}) \approx & \frac{1}{2}\left[\int_{\bm{q}} P^L_{q}P^L_{|\bm{k}-\bm{q}|} \frac{\bm{k}\cdot(\bm{k}-\bm{q})}{|\bm{k}-\bm{q}|^2}F_{ZA}(\bm{q},\bm{k}-\bm{q}) \right]^2 \\
            & + \frac{1}{2} \int_{\bm{q}_1} P^L_{q_1}P^L_{|\bm{k}-\bm{q}_1|} \frac{\bm{k}\cdot(\bm{k}-\bm{q}_1)}{|\bm{k}-\bm{q}_1|^2} \frac{\bm{k}\cdot\bm{q}_1}{q_1^2} \int_{\bm{q}_2} P^L_{q_2} P^L_{|\bm{k}-\bm{q}_2|} F_{ZA}^2(\bm{q}_2,\bm{k}-\bm{q}_2).
    \end{split}
\end{equation}

\section{EUCLID Forecast Parameters}

The parameters used to estimate the measurement errors on EUCLID spectra were obtained from \cite{euclid_2018_amendola}. We have considered the volume of a redshift slice ranging from $z=0.85$ to $z=0.95$, spanning an observed area of $15.000 \text{deg}^2$ and using the cosmology from \cite{planck2018} to compute the comoving distances.

\begin{table}[h]
\centering
\bgroup
\def\arraystretch{1.5}%  1 is the default, change whatever you need
\begin{tabular}{|c|c|c|c|}
\hline
$V_E[$Gpc$/h]^3$ & $\bar{n}_{\text{ELG}}[h/$Mpc$]^{-3}$ & $z$ & $b_1^L$\\
$33.05$ & $1.83\times 10^{-3}$ & $0.85 - 0.95$ & $0.38$ \\
\hline
\end{tabular}
\egroup
\caption{ Fiducial parameters used to estimate the measurement error for galaxy power-spectra in an EUCLID-like survey. }
\label{Tab:pars_euclid}
\end{table}

\begin{table}[h]
\centering
\bgroup
\def\arraystretch{1.5}%  1 is the default, change whatever you need
\begin{tabular}{|c|c|c|c|c|}
\hline
 & $b^L_1$ & $b^L_2$ & $b_{s^2}^L$ & $b_{\nabla^2\delta}^L$\\
Fiducial Case & $0.38$ & $-7.16\times 10^{-4}$ & $-5.04\times 10^{-3}$ & $4.01\times 10^{-2}$\\
Worst Case & $0.38$ & $0.80$ & $1.49$ & $4.01\times 10^{-2}$\\
\hline
\end{tabular}
\egroup
\caption{ Bias parameters used to infer the minimum simulation volume necessary to have the model error under control. We quote values both for a fiducial case, in which we choose the values preferred by the coevolution relations obtained in \cite{zennaro2021priors}, and a worst case scenario, in which we choose the largest values allowed by this analysis for $b^L_2$ and $b^L_{s^2}$. }
\label{Tab:pars_euclid}
\end{table}

\section{Details of Optimization}
\label{section:optimization_details}

\begin{figure}[h]
    \centering
    \includegraphics[width=\textwidth]{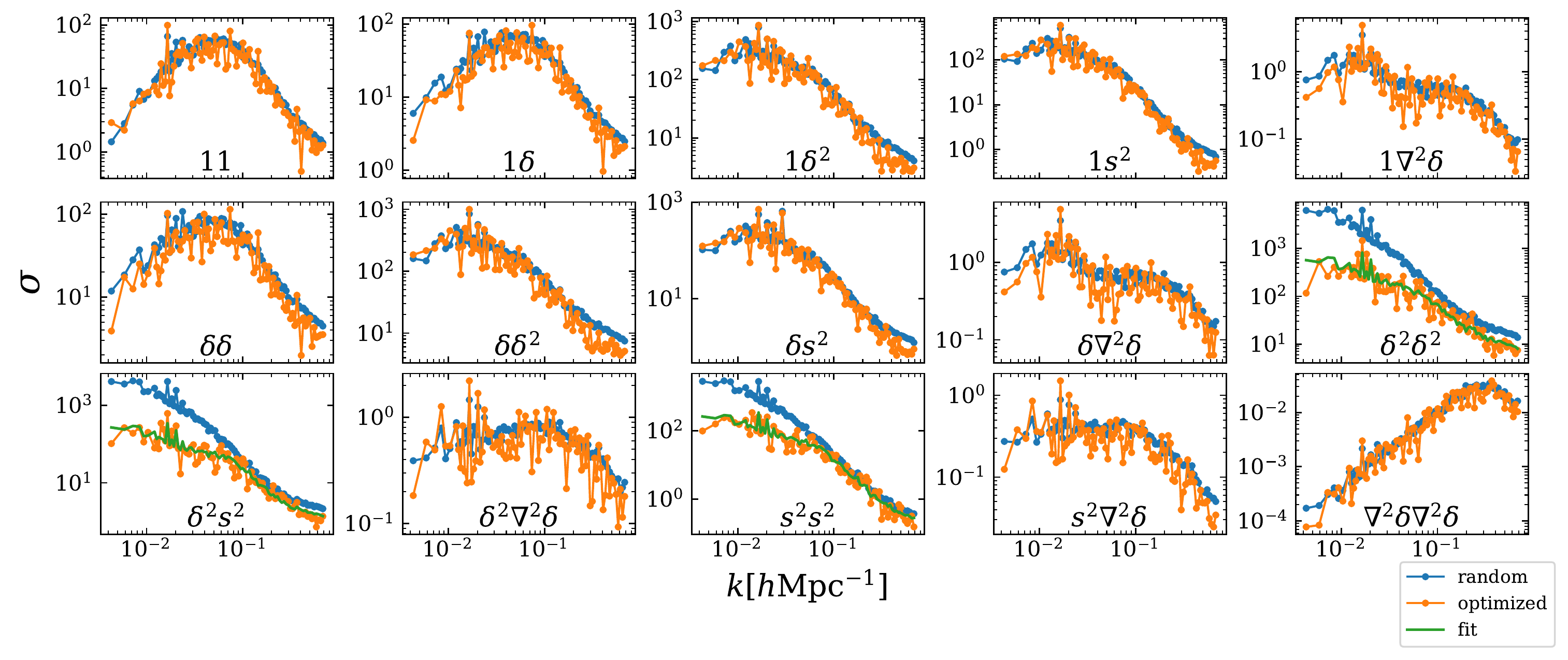}
    \caption{This figure shows a comparison between the variances for usual F\&P simulations, and those initiliazed with the optimal phase-fields derived in section \ref{section:ph_field}. The green lines show the fits made to reproduce the new values of the variances.}
    \label{fig:var_optimal_comp_fits}
\end{figure}

In section \ref{section:ph_field} we have derived optimal phase-fields which can be used to initialize simulations and obtain spectra with reduced variances. Figure (\ref{fig:var_optimal_comp_fits}) shows a comparison between the variances computed from the 200 usual F\&P simulations, and those computed from a set of 5 F\&P simulations initialized with distinct realizations of the optimal phase-fields. With the purpose of including the effects of this suppression into the calculation of $\sigma_{gg}$ -- see equation (\ref{eq:bias_exp_var}) -- we fitted the ratio $\sigma^{op}/\sigma^{F\&P}$ using the following expression
\begin{equation}
    f(k,a,b,c,d,e) = a\left(b + c\tanh{\left[(k - d)e\right]} \right).
\end{equation}
Using this fit we could then approximate the optimized variance for any value of the simulated volume, cosmology or redshift using
\begin{equation}
    \sigma^{op}(k, \Theta, z, V) = \sigma^{F\&P}(k, \Theta, z, V) f(k, a_0, b_0, c_0, d_0, e_0),
\end{equation}
in which we have represented the cosmological parameters by $\Theta$, $z$ is the redshift, $V$ is the simulated volume, and $a_0, b_0, \cdots$ are the fitted values of the parameters.

\printbibliography

\acknowledgments

The authors acknowledge the support of the ERC-StG number
716151 (BACCO). FM acknowledges partial support from FAPESP via the fellowship 2019/01631-0. The authors acknowledge useful discussions with C. Rampf and R. Voivodic. We also thank O. Hahn, C. Rampf and R. Voivodic for valuable suggestions and comments on the draft.

\end{document}